\documentclass[twocolumn,twocolappendix]{aastex631}
\usepackage{comment}
\usepackage{chemformula}
\usepackage{booktabs}
\usepackage{rotating}
\usepackage{mathtools}
\usepackage{graphicx}
\usepackage{soul}
\usepackage{hyperref}

\shorttitle{Binding Energies of S-bearing Species}
\shortauthors{J. Perrero et al.}

\begin{document}

\title{Binding Energies of Interstellar Relevant S-bearing Species on Water Ice Mantles: A Quantum Mechanical Investigation}

\correspondingauthor{Albert Rimola, Piero Ugliengo}

\author[0000-0003-2161-9120]{Jessica Perrero}
\affiliation{Departament de Qu\'{i}mica, Universitat Aut\`{o}noma de Barcelona, Bellaterra, 08193, Catalonia, Spain}
\affiliation{Dipartimento di Chimica and Nanostructured Interfaces and Surfaces (NIS) Centre, Universit\`{a} degli Studi di Torino, via P. Giuria 7, 10125, Torino, Italy}

\author[0000-0002-2147-7735]{Joan Enrique-Romero}
\affiliation{Departament de Qu\'{i}mica, Universitat Aut\`{o}noma de Barcelona, Bellaterra, 08193, Catalonia, Spain}

\author[0000-0001-7819-7657]{Stefano Ferrero}
\affiliation{Departament de Qu\'{i}mica, Universitat Aut\`{o}noma de Barcelona, Bellaterra, 08193, Catalonia, Spain}

\author[0000-0001-9664-6292]{Cecilia Ceccarelli}
\affiliation{Univ. Grenoble Alpes, CNRS, Institut de Plan\'{e}tologie et d'Astrophysique de Grenoble (IPAG), 38000 Grenoble, France}

\author[0000-0003-2733-5372]{Linda Podio}
\affiliation{INAF, Osservatorio Astrofisico di Arcetri, Largo E. Fermi 5, 50125 Firenze, Italy}

\author[0000-0003-1514-3074]{Claudio Codella}
\affiliation{Univ. Grenoble Alpes, CNRS, Institut de Plan\'{e}tologie et d'Astrophysique de Grenoble (IPAG), 38000 Grenoble, France}
\affiliation{INAF, Osservatorio Astrofisico di Arcetri, Largo E. Fermi 5, 50125 Firenze, Italy}

\author[0000-0002-9637-4554]{Albert Rimola}
\affiliation{Departament de Qu\'{i}mica, Universitat Aut\`{o}noma de Barcelona, Bellaterra, 08193, Catalonia, Spain}
\email{albert.rimola@uab.cat}

\author[0000-0001-8886-9832]{Piero Ugliengo}
\affiliation{Dipartimento di Chimica and Nanostructured Interfaces and Surfaces (NIS) Centre, Universit\`{a} degli Studi di Torino, via P. Giuria 7, 10125, Torino, Italy}
\email{piero.ugliengo@unito.it}

\begin{abstract}

Binding energies (BEs) are one of the most important parameters for astrochemical modeling determining, because they govern whether a species stays in the gas-phase or is frozen on the grain surfaces. It is currently known that, in the denser and colder regions of the interstellar medium, sulphur is severely depleted in the gas phase. It has been suggested that it may be locked into the grain icy mantles. However, which are the main sulphur carriers is still a matter of debate. This work aims at establishing accurate BEs of 17 sulphur-containing species on two validated water ice structural models, the proton-ordered crystalline (010) surface and an amorphous water ice surface. We adopted Density Functional Theory (DFT)-based methods (the hybrid B3LYP-D3(BJ) and the hybrid meta-GGA M06-2X functionals) to predict structures and energetics of the adsorption complexes. London's dispersion interactions are shown to be crucial for an accurate estimate of the BEs due to the presence of the high polarizable sulphur element. While on the crystalline model the adsorption is restricted to a very limited number of binding sites with single valued BEs, on the amorphous model several adsorption structures are predicted, giving a BE distribution for each species. With the exception of few cases, both experimental and other computational data are in agreement with our calculated BE values. A final discussion on how useful the computed BEs are with respect to the snow lines of the same species in protoplanetary disks is provided.

\end{abstract}

\keywords{ Unified Astronomy Thesaurus concepts: Surface ices (2117) --- Interstellar dust (836) --- Interstellar molecules (849) --- Dense interstellar clouds (371) --- Interstellar medium (847) --- Solid matter physics (2090) --- Interstellar dust processes (838) --- Computational methods (1965)}

\section{Introduction} \label{sec:intro}

Stars like our Sun begin their journey in the interstellar medium (ISM), where star formation takes place in dense (10$^3$--10$^4$ cm$^{-3}$) and cold (10 K) regions, the so-called molecular clouds. In the past, it was thought that the ISM was a too harsh environment for molecules to survive and thrive. However, almost eighty years ago, the first diatomic molecules (the  methylidyne radical CH, its cation CH$^+$, and the cyano radical CN) were detected by means of optical and ultraviolet spectroscopy \citep{swings:1937,mckellar:1940,douglas:1942}, which indeed indicated that chemistry is present and has an important role in the ISM.

At present, we count more than 270 detected gas-phase species \citep{mcguire:2021}, a number which constantly increases given the enhanced performances of the observational facilities (e.g., Yebes 40-m, IRAM 30-m, NOEMA, ALMA and GBT), and the expectation of the new results provided by the James Webb Space Telescope (JWST). As a consequence of these thrilling discoveries, curiosity concerning the chemistry of the ISM has been growing more and more keen.

The characteristic densities and temperatures of molecular clouds allow (i) the molecules to survive, since the external UV fields are highly diminished, and (ii) the formation of thick icy mantles coating the surfaces of dust grains, due to the adsorption and in-situ synthesis of species and their subsequent hydrogenation. The resulting icy mantles are dominated by water, but they also contain other volatile species like CO, CO$_2$, CH$_3$OH and NH$_3$, and hence are referred to as ``dirty ices'' \citep{boogert:2015}.

The phenomena occurring on the icy mantle surfaces are of paramount importance from the chemical point of view.
Indeed, ice mantles are thought to facilitate the occurrence of reactions taking place at their surfaces, exhibiting mainly three functions: i) as pre-concentrators of chemical species, especially relevant in the low-density environments of molecular clouds, where the rate of collision between gas-phase species is extremely small, ii) as chemical catalysts, by decreasing the activation energies of the reactions, and hence overcoming the energy barriers under the interstellar conditions \citep{Zamirri:2019c}, and iii) as third bodies, by absorbing the large energy excess released by exothermic reactions, without undermining the stability of the newly formed products \citep{pantaleone:2020,pantaleone2021}. The first and the third points are the keys to successful recombination reactions. 

For a grain surface reaction to take place, at least one of the two reactants needs to be adsorbed on the grain. The adsorption is regulated by two variables: the dust temperature, and the binding energy (BE) of the adsorbed species. The latter quantity defines how strong the interaction between the species and the surface is, and dictates the chance that a given adsorbed species can be ejected into the gas phase: the higher the BE, the higher the temperature required to thermally desorb. In the current astrochemical models, BE values are important input parameters, not only to consider likely desorption events \citep[e.g.][]{penteado:2017} but to simulate diffusion processes on the surfaces, as the diffusion barriers are usually assumed to be a fraction of the BE \citep{mispelaer:2013, karssemeijer:2014, lauck:2015, ghesquiere:2015, he:2017, he:2018, cuppen:2017, cooke:2018, mate:2020, kouchi:2020}.
Therefore, BEs are crucial parameters in determining the ISM chemistry and its resulting composition \citep{penteado:2017, wakelam:2017, ferrero:2020, enrique-romero:2021}.

\subsection{Previous studies of the BEs}\label{subsec:intro_be}

Determining BEs is a challenging task \citep{Minissale:2022}. From an experimental point of view, effective and coverage-dependent BEs are obtained via temperature programmed desorption (TPD) experiments. TPD consists of two steps: first, the substrate, maintained at a constant temperature, is exposed to the adsorption of the species; then, the temperature is increased until the desorption of the species, which are collected and analyzed by a mass spectrometer. The BEs are then extracted by applying the direct inversion method through the Polanyi–Wigner equation \citep[e.g.,][]{dohnalek:2001}. 
Despite its great usefulness, this technique shows two limitations: i) the obtained BEs depend not only on the morphology and the composition of the substrate but also on the regimes in which the experiments are performed, i.e., sub-monolayer, monolayer or multilayer \citep[e.g.,][]{noble:2012, he:2016, chaabouni:2018} and will return coverage-dependent BEs, and ii) it measures a desorption enthalpy, a quantity which is equal to the BE only in the absence of other activated processes, like ice restructuring, usually assumed to be negligible \citep{he:2016}. 

Despite the numerous literature studies that have investigated the desorption processes by means of TPD \citep[e.g.,][]{collings:2004, noble:2012, dulieu:2013, he:2016, fayolle:2016, smith:2016}, only a limited variety of molecules has been considered, in particular stable closed-shell species, in contrast to the large variety of interstellar species. Moreover, the employed substrates are almost invariably  water ices.

A possible alternative to TPD is to adopt state-of-the-art computational chemistry to simulate the adsorption process. The simulation can handle both closed- and open-shell species, putting some extra care for the latter. The main limitation is the trade-off between the system size (the icy grain model) and the accuracy of the calculation.  
A number of computational studies have reported the BEs of important astrochemical species like H, H$_2$, N, O, HF, CO, and CO$_2$ on icy surfaces modelled by periodic/cluster crystalline/amorphous systems \citep[e.g.][]{alhalabi:2007, karssemeijer:2014, zamirri:2017, shimonishi:2018, zamirri:2019, bovolenta:2020, ferrero:2020}. 
Other works addressed the BEs of a large number of species, adopting a very approximated models for the substrate. Two examples are the works of \cite{wakelam:2017} and \cite{das:2018}, in which the ice surface simulated by water clusters of minimal nuclearity allowed the prediction of the BEs for a hundred of species. In \cite{wakelam:2017}, a single water molecule was adopted to represent the ice surface, while recovering the missing components of the BEs through a clever fitting procedure against selected experimental BE values. In \cite{das:2018}, the adoption of clusters from 1 to 6 water molecules allowed to study the BE of 16 species, with the puzzling result showing water tetramer as the best model for ice.

One way to generate computer models for amorphous ices is by using water clusters cut out from a crystalline ice,  heating them up by molecular dynamics runs and by quenching at low temperatures \citep{shimonishi:2018,Rimola-Minerals}.
Another approach is to refer to a model repeated in space by the periodic boundary conditions. This allows to simulate both the crystalline ice (here adopting an already studied proton ordered model by \cite{casassa:1997}) and an amorphous one. In the latter case, the usual procedure is starting from a large unit cell of a crystalline ice and performing a number of heating/freezing cycles to arrive to an amorphous ice model. 
Another strategy, recently proposed by the ACO-FROST scheme \citep{germain2022}, consists in growing icy cluster by a random step-by-step addition of molecular water molecules.

More recently, \cite{ferrero:2020} computed a large set of BEs for 21 astrochemically-relevant species, on both a crystalline and an amorphous surfaces of water ice. Both models ensured that hydrogen-bond cooperativity (of fundamental relevance to obtain accurate BE values) is fully taken into account, at variance with the small ice clusters adopted by \cite{wakelam:2017} and \cite{das:2018}. For the crystalline icy model, only very few adsorption sites were available, limiting the BEs to one or two distinct values per species. For the amorphous ice model, however, the complex surface morphology allowed the calculation of multiple BE values for each species, resulting in a BE range better describing the variety of binding sites expected on real amorphous ice mantles.

In the present work, we apply the methodology of \cite{ferrero:2020} in order to enlarge the set of BEs. The focus here is on the adsorption of 17 interstellar relevant S-bearing species, namely: CS, C$_2$S$^{\bullet\bullet}$, C$_3$S, C$_4$S$^{\bullet\bullet}$, CH$_3$SH, H$_2$CS, HS$^{\bullet}$, H$_2$S, HS$_2^{\bullet}$, H$_2$S$_2$, NS$^{\bullet}$, OCS, S$^{\bullet\bullet}$, S$_2^{\bullet\bullet}$, SO$^{\bullet\bullet}$, SO$_2$, (dots indicating the unpaired electrons). The chosen species satisfy two criteria: i) they are neutral species, since positive ions become neutral when landing on negatively charged dust grains \citep{Walmsley:2004, Ceccarelli:2005, Rimola:2021} as long recognized \citep{Draine:1987, Rae:2004, Mason:2014}, and ii) they envisage at most 6 atoms to ensure relatively compact structures needed to properly probe the largest number of adsorption sites on ice models with relatively small unit cells.

\subsection{The sulphur depletion} \label{subsec:sulphur}

The choice of studying the BEs of S-bearing species is due to a long standing issue in the field of Astrochemistry: the sulphur depletion problem. In dense clouds, sulphur is severely depleted from the gas phase \citep[e.g.][]{tieftrunk:1994,Ruffle:1999,wakelam:2004,phuong:2018,vastel:2018,vanthoff:2020} by more than two orders of magnitude with respect to the Solar System S abundance ([S]/[H] = 1.8x$10^{-5}$, \cite{anders1989,woods:2015}). Therefore, sulphur may freeze out on dust grains after its hydrogenation, or react with other species in a similar way as atomic oxygen. H$_2$S was expected to be the main sink of sulphur on dust grains \citep{garrod:2007, escobar:2011}, but it has never been directly detected on interstellar ices yet. The only species detected in the ice mantles are OCS and, tentatively, SO$_2$ \citep{boogert:2015}. Accordingly, the most important S-carriers are still unknown. Nowadays, the most likely species that act as reservoirs of S are thought to be organo-sulphur compounds \citep{laas:2019} and polysulphanes (H$_2$S$_n$) \citep{druard:2012}. These are refractory species that, once trapped on the mantles or in the core of the grains, cannot desorb and, therefore, become undetectable \citep{woods:2015}. In the last three years, seven new gas-phase S-bearing species were identified, most of them in TMC-1 \citep{cernicharo:2021, rodriguez:2021}, and characterized by a C-S chemical bond. 

A recent review of the laboratory experiments performed on the chemistry of sulphur in the condensed phase has highlighted that the formation of SO$_2$, SO$_3$, hydrates of H$_2$SO$_4$ and related species is very common when performing photolysis, proton-irradiation and radiolysis of mixed ices. However, neutral-neutral reactions have yet to be deeply explored. Moreover, gas phase chemistry is likely to be a major contributor in sulphur astrochemistry \citep{mifsud:2021}.

The formation of organosulphur molecules was observed experimentally for the first time in the work of \cite{ruf:2019}, where a 2:1:1 mixture of H$_2$O:CH$_3$OH:NH$_3$ ice was bombarded with S$^{7+}$ ions, although showing a possible source of compounds relative to more evolved environments, like icy moons, Kuiper belt objects, comets and their building blocks.

In diffuse clouds, sulphur is mostly present in its ionized atomic form \ch{S+} and close to the cosmic abundance \citep{jenkins:2009}, it progressively depletes as the cloud evolves \citep{hilyblant:2022}, while sulphur depletion in protoplanetary disks and later stages is explained by the presence of refractory sulphide minerals, such as FeS, in the grains \citep{keller:2002,kama:2019}. 

With the present work, thus, we aim to contribute to this puzzling subject by providing a well-suit of accurate BE values of several S-bearing species, which can be used in numerical modelling studies aimed at rationalising observations and laboratory experiments.

\section{Methodology} \label{sec:computational}

Since the calculations follow the same procedure as \cite{ferrero:2020}, we refer to this work for the equations used here to compute the BEs of the S-bearing species set. Moreover, the reader can find details on these  expressions in the Appendix (\S ~\ref{sec:appendix}). A thorough guide to the computation of BEs is available in the Supplementary Material.

\subsection{Computational Details} \label{subsec:method}
We adopted a periodic approach to model the water ice surfaces and the adsorption of the 17 S-bearing species on them. The calculations were performed with the periodic \textit{ab initio} code \textsc{crystal17} developed by \cite{crystal}, which adopts localized Gaussian functions as basis sets. This code exploits the local combination of atomic orbitals (LCAO) approximation, expanding the Bloch functions of periodic systems as a linear combination of atom-centered Gaussian functions. This software implements both the Hartree-Fock and Kohn-Sham self-consistent field methods to solve the electronic Schr\"odinger equation, taking advantage of the symmetry of the system, when present. Thanks to its capability to simulate systems with periodicity ranging from zero (molecules) to three (solid bulks) dimensions, it allows to rigorously define true slabs to modelling surfaces without the need of artificial replica along the direction perpendicular to the slab plane, as usually implemented in codes adopting plane waves basis sets. 

On the crystalline model, the hybrid B3LYP   \citep{lyp:1988,becke:1988,becke:1993} and the hybrid meta-GGA M06-2X \citep{zhao:2008} density functionals were used in the geometry optimizations and in computing the BEs of closed-shell and open-shell species, respectively. To account for dispersion interactions, the hybrid B3LYP functional was combined with the Grimme's D3 empirical correction with the Becke-Johnson (BJ) damping scheme (i.e., B3LYP-D3(BJ) \citep{sure:2013,grimme:2010,grimme:2011}. The hybrid meta-GGA M06-2X functional was used together with a spin-unrestricted formalism \citep{pople:1995} to model the adsorption of open-shell species (with one or two unpaired electrons) because it better describes their adsorption properties \citep{ferrero:2020} due to its higher percentage of exact exchange (54\% compared to the 20\% of B3LYP). Note that for M06-2X, the D3(BJ) correction for dispersion was not applied because, for this functional, the description of the dispersion component is already included in its definition. For this reason, the contribution of the dispersion component to the BEs can be worked out only for the closed-shell species (computed at B3LYP-D3(BJ)). The two DFT methods have been used in combination with the Ahlrichs triple zeta valence quality basis set supplemented with a double set of polarization functions (i.e., A-VTZ*, \cite{schafer:1992}). Thus, for the crystalline models, the BEs were computed from the DFT energies that resulted from the DFT optimized structures. This scheme is hereafter referred to as DFT//DFT. All the BEs were corrected for the basis set superposition error (BSSE) through the counterpoise correction method by \cite{boysb:1970}. 

Finally, in order to check for the accuracy of the DFT//DFT BEs and without the aim to perform a benchmarking study, a refinement at CCSD(T) level was performed by applying the ONIOM2 scheme as proposed by \cite{dapprich:1999} to a number of species representative of the set. Given the dependency of CCSD(T) method on the basis set, the \cite{dunning} correlation consistent family of \textit{cc-pVnZ} basis sets (where \textit{n} is 2, 3, 4, 5, ... and corresponds to a $\zeta$ of D,T,Q,5,...) was employed. This allows us to perform an extrapolation of the desired property, here the BE, by plotting it against $1/n^3$. We checked that the numbers so obtained approached the values given by DFT//DFT. The Jun-cc-pVnZ basis set (with n = 2, 3 and 4) was chosen to perform the calculations, with the exception of SO$_2$, for which the aug-cc-pVnZ basis set (where one more diffuse function is added to all the atoms, including H) was adopted to recover the whole interaction between SO$_2$ and the two water molecules of the \textit{model} system.
See \S ~\ref{sec:appendix} for more details on the BSSE correction and on the ONIOM2 scheme.

On the amorphous ice model, the DFT//DFT scheme can hardly be carried out due to the larger unit cell size compared with the crystalline one. Therefore, we  adopted the computationally cheaper semi-empirical HF-3c method for the adsorption on the amorphous ice surface model. HF-3c is a Hartree-Fock (HF)-based method adopting a minimal basis set (MINI-1, \citep{tatewaki:1980}), to which three empirical corrections (3c) are added \citep{sure:2013}: i) the dispersion energy (D3(BJ)) for non-covalent interactions \citep{grimme:2010}; ii) a short range bond correction (SRB) to recover the systematically overestimated covalent bond lengths for electronegative elements, due to the MINI-1 basis set \citep{grimme:2011,brandenburg:2013}; and iii) the geometrical counterpoise (gCP) method developed by Grimme \citep{kruse:2012} to  \textit{a priori} remove the BSSE. On the optimized HF-3c geometries, DFT single point energy calculations were performed to compute the BEs (this procedure hereafter referred to as DFT//HF-3c), which were also corrected for BSSE. Before adopting DFT//HF-3c for the amorphous system we proved its accuracy compared to DFT//DFT for the crystalline ice, resulting in an excellent match. 

For both the crystalline and amorphous systems, geometry optimizations were carried out by relaxing both the internal atomic positions inside the unit cell and the cell parameters.

Each stationary point was characterized by means of harmonic frequency calculations at the $\Gamma$ point by diagonalizing the mass-weighted Hessian matrix of the second-order energy derivatives with respect to atomic displacements \citep{pascale:2004,zicovich:2004}. Each Hessian matrix element was computed numerically by means of a 6-point formula based on two displacements of $\pm$ 0.003 \r{A} from the minimum along each Cartesian coordinate. For the crystalline ice, computed at the DFT level, only a portion of the system constituted by the adsorbed species plus the two closest water molecules to the adsorbate was considered to build the Hessian matrix. In the case of the amorphous ice, the entire system was considered, and the full Hessian matrix was computed at the HF-3c level. For this later case, however, the frequencies were only used to check that the complexes are actual minima of the potential energy surfaces (all the frequencies are real), since they are expected to not be accurate enough because of the approximated HF-3c methodology. In contrast, from the DFT frequencies computed for the fragments, we computed the zero-point vibrational energy (ZPE) corrections to: i) correct the electronic BEs for the crystalline ice by including the corresponding ZPE contributions; and ii) derive a scaling factor to apply to the BEs computed on the amorphous ice model to arrive at ZPE-corrected values  without the need to explicitly run a full DFT frequency calculation (see \S ~\ref{subsec:be} for more details on these two points).

\subsection{Ice Models} \label{subsec:ice}

\begin{figure}
    \centering
    \includegraphics[width=\linewidth]{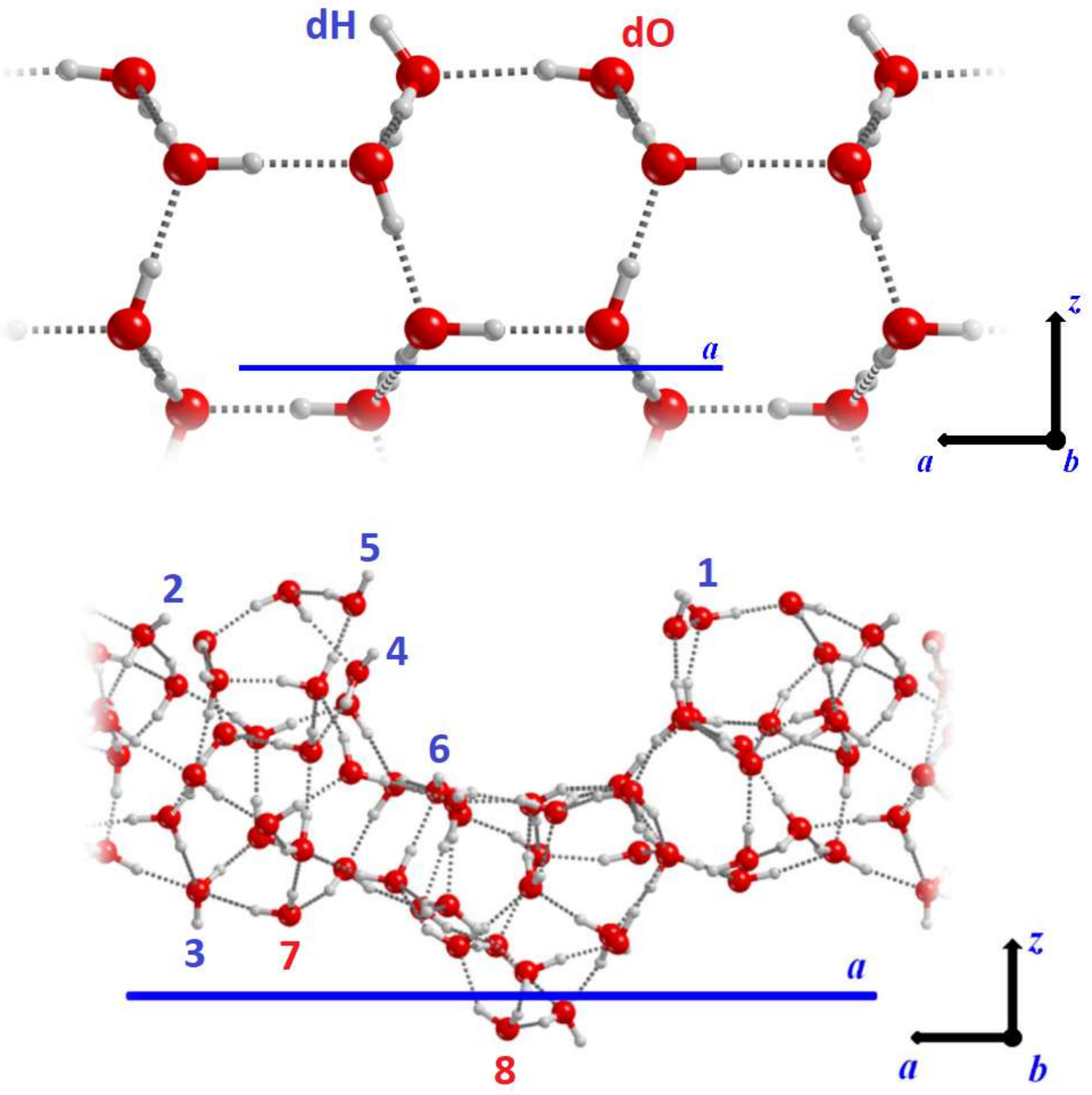}
    \caption{Top panel: side view of the crystalline (010) water P-ice slab model (along the \textit{b} lattice vector). Bottom panel: side view of the amorphous water ice slab model (along the \textit{b} lattice vector). Numbers identify the different adsorption sites: those in blue represent dangling hydrogen atoms (dH), while those in red represent dangling oxygen atoms (dO).}
    \label{fig:ice_model}
\end{figure}

Interstellar ices are thought to be mostly formed by amorphous solid water \citep{watanabe:2008,boogert:2015}. However, as a starting point of our study, we have chosen a crystalline model (see top panel of Figure \ref{fig:ice_model}). This choice is due to two main reasons: (i) crystalline structures are well defined due to symmetry constraints and are computationally cheap; and (ii) no definite structure for amorphous ice models are available. Moreover, regions rich in crystalline ices have been observed in protoplanetary disks and stellar outflows \citep{molinari:1999,terada:2012}. Our ice surface model derives from the bulk of the proton-ordered P-ice, which was cut along the (010) surface defining a 2D periodic slab model \citep{casassa:1997,zamirri:2018}. The thickness of the ice (10.9 \r{A}) was enough to converge  the corresponding surface energy. We adopted a 2 $\times$ 1 supercell slab model, consisting of twelve atomic layers and cell parameters of $\left| a \right|$ = 8.980 \r{A} and $\left| b \right|$ = 7.082 \r{A} (at B3LYP-D3(BJ)/A-VTZ* level), which are large enough to avoid lateral interactions between the adsorbed molecules in different replicas of the ice cell, thereby simulating the adsorption of isolated species. The supercell shows two dangling hydrogen (dH) atoms and two dangling oxygen (dO) atoms as binding sites (top panel of Figure \ref{fig:ice_model}). The structure of the ice is such as to ensure a null electric dipole along the non-periodic z-axis. This is a direct consequence of the symmetry of the system, which shows two identical faces both at the top and at the bottom of the model. Therefore, the adsorption was modeled only on the top face of the system.

The more realistic amorphous model, already used in \cite{ferrero:2020} (see bottom panel of Figure \ref{fig:ice_model}) consists of 60 water molecules per unit cell. The cell parameters are $\left| a \right|$ = 20.355 \r{A}, $\left| b \right|$ = 10.028 \r{A} and $\left| \gamma \right|$ = 102.983° (at HF-3c). The different structural features between the upper and lower surfaces are responsible for the presence of a small electric dipole moment across the non-periodic direction, and also for the presence of a variety of different binding sites. Therefore, to model the adsorption of the S-bearing species, we selected eight characteristic adsorption sites (highlighted in the bottom panel of Figure \ref{fig:ice_model}). The sites are chosen as follows: cases 1, 2 and 5 are located on the top-surface of the ice, case 4 and 6 are inside the cavity of the model, and case 3, 7 and 8 are on the bottom-surface of the model. Binding sites from 1 to 6 are dH, while 7 and 8 are dO.

\section{Results} \label{sec:results}

 \begin{figure}
    \centering
    \includegraphics[width=\linewidth]{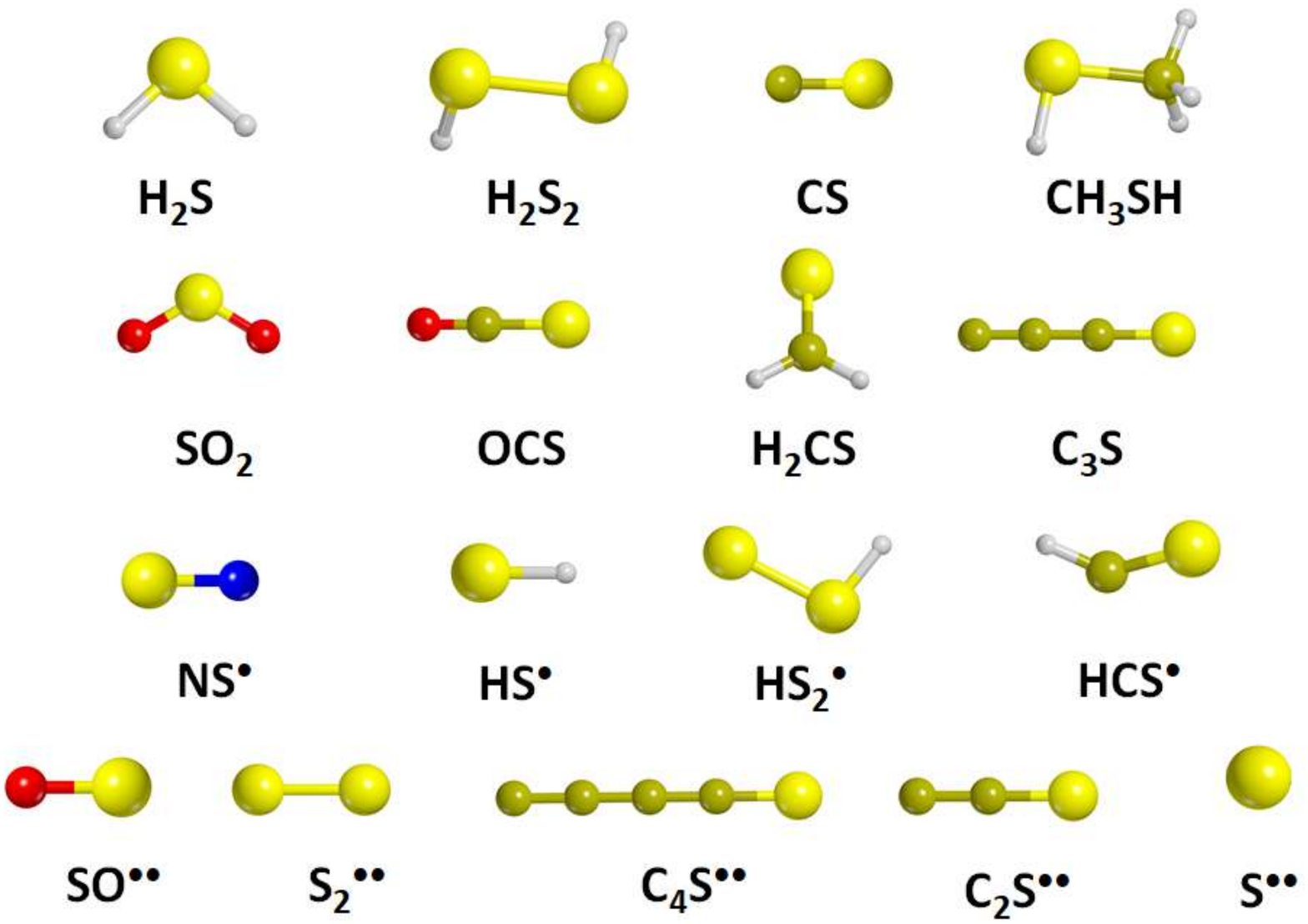}
    \caption{Set of 17 S-bearing species considered in this work. Dots represent unpaired electrons.}
    \label{fig:species}
\end{figure}

In this work, the adsorption of 17 S-bearing species (shown in Figure \ref{fig:species}) has been simulated, of which eight are closed-shell species and nine open-shell ones (with unpaired electrons as indicated by the dots in Figure \ref{fig:species}). For the open-shell species, the electronic ground state of each one was checked and found to be in agreement with those reported by \cite{Woon:2009} and \cite{das:2018}. Therefore, NS$^{\bullet}$, HS$^{\bullet}$, HS$_2^{\bullet}$ and HCS$^{\bullet}$ are doublets (with one unpaired electron) and S$^{\bullet\bullet}$, SO$^{\bullet\bullet}$, S$_2^{\bullet\bullet}$, C$_2$S$^{\bullet\bullet}$ and C$_4$S$^{\bullet\bullet}$ are triplets (with two unpaired electrons). Each species was placed manually on the ice surface models according to the molecule/surface electrostatic potential complementarity, and from these manually built systems, geometry optimizations were performed.

\subsection{BEs on Crystalline Ice} \label{subsec:be_cry}

\subsubsection{DFT//DFT BEs} \label{subsec:dft}

Table \ref{tab:BE_cry} reports the computed BE values as described in the computational section.

The role of the dispersion interactions to the BEs was assessed by considering the D3(BJ) contribution only to the BE. In other words, for each full B3LYP-D3(BJ) optimized structure, we split the B3LYP-D3(BJ) total energy in a pure electronic DFT term coming from the B3LYP and the pure dispersive D3(BJ) one. This approach maximizes the weight of the dispersion because the attractive nature of the D3(BJ) term brings the adsorbates as close as possible to the icy surface which, in turn, boost the D3(BJ) value. A different approach would be to optimize the geometries at B3LYP only and then evaluate the D3(BJ) term through a single point energy evaluation (referred to as B3LYP-D3(BJ)//B3LYP scheme). The missing D3(BJ) term during the geometry optimization causes both an elongation of the average distance and important changes in the spatial arrangement of the adsorbate compared to the full B3LYP-D3(BJ) cases. Therefore, the dispersion contribution for the B3LYP-D3(BJ)//B3LYP is expected to be smaller (underestimated) than the one from the full B3LYP-D3(BJ) optimization calculation. We checked both approaches (data not reported) for two set of molecules, O-bearing (CO, H$_2$O, CO$_2$, CH$_3$OH, H$_2$CO) and S-bearing (CS, H$_2$S, OCS, CH$_3$SH, H$_2$CS) adsorbed on the P-ice crystalline surface. We found the B3LYP structures involving CS and, especially OCS, to be rather different from the corresponding B3LYP-D3(BJ) ones. The same was found for CO and CH3OH cases. Nonetheless, the percentage average difference in the dispersion between the two approaches is a modest 11\% in favor of the B3LYP-D3(BJ), as expected. In this work, we were interested in establishing the relative weight of dispersion (\textit{rwd}) between S and O-containing species, rather than its the absolute value. For the cases described above, we computed a \textit{rwd} percentage average difference between B3LYP-D3(BJ)//B3LYP and full-optimized B3LYP-D3(BJ) of only 4\%, without any order reverting in the dispersion weight. Therefore, in agreement with this test case, the data shown in this work are relative to the full B3LYP-D3(BJ) approach, aware that the role of dispersion may be overestimated.

At DFT//DFT level, computed \textit{BE disp} values span a range from 2400 K to 6900 K, depending on the kind of interaction that the adsorbate experiences with the ice surface, i.e., hydrogen bonds, interactions between both permanent and instantaneous dipole moments, and dispersive interactions.
According to that, we can distribute the molecules in four different groups:

\begin{description}
\item[Molecules of Group I] (CH$_3$SH, H$_2$CS, H$_2$S, H$_2$S$_2$, HCS$^{\bullet}$, HS$^{\bullet}$, and HS$_2^{\bullet}$): these molecules contain groups that can act as H-bond donors and H-bond acceptors. It is worth noting that in the cases of H$_2$CS and HCS$^{\bullet}$ the H-bond donor groups are weak.
\item[Molecules of Group II] (OCS, SO$_2$, NS$^{\bullet}$, and SO$^{\bullet\bullet}$): these molecules feature only H-donor acceptor groups because they contain atoms that are more electronegative than S.
\item[Molecules of Group III] (CS, C$_2$S$^{\bullet\bullet}$, C$_3$S, and C$_4$S$^{\bullet\bullet}$): molecules that can establish one H-bond interaction with their C-end atom and a surface dH.
\item[Molecules of Group IV] (S$^{\bullet\bullet}$ and S$_2^{\bullet\bullet}$): species consisting of only S atoms so that dispersive interactions are expected to be the main binding driving forces. 
\end{description}

The molecule with the smallest dispersion contribution to its BE is H$_2$S (of Group I), which, like water, is capable to act as both H-bond donor and acceptor, hence involving two water molecules of the ice surface in its adsorption. H$_2$S, H$_2$S$_2$, H$_2$CS and CH$_3$SH (of the same Group I) show this same adsorption feature. However, the presence of a S-S or a C-S bond enhances the contribution of dispersion forces, that can reach up to ca. 70\% of the BE, as in H$_2$S$_2$. The open-shell species HS$^{\bullet}$ and HS$_2^{\bullet}$  (also of Group I), have smaller BEs than their hydrogenated counterparts due to the absence of one H atom. 

For molecules of Group II, the sulphur atom interacts through dispersive forces with dO of the nearest surface water molecules. SO$_2$, thanks to its two oxygen atoms that act as H-bond acceptors, presents the highest BE of the entired set of tested molecules (6880 K). OCS, which shows an almost null dipole moment, presents the largest dispersion contribution to the BE of the entire set ($>$90\%), hence being one of the species less bound to the ice.

Molecules of Group III form indeed a H-bond between their C-ends and a surface dH. This interaction, moreover, is enhanced due to the dipole moment of the C chains and to the carbene-like character of the C-end atom. Furthermore, these species can bind strongly to the surface thanks to the onset of dispersive forces that become stronger with the size of the molecule. Despite these premises, the BEs of these species are similar (4000-4500 K), with the exception of C$_3$S, which has a higher BE (6723 K at the most) because of its adsorption geometry. That is, once C$_3$S adsorbs on the surface, it is arranged in rows with a distance of only 3.1 \r{A} between the S-end and the C-end of two neighbouring molecules, therefore maximizing  favourable head-to-tail interactions between dipoles. This results in an energetic gain due to the lateral interactions between adsorbate molecules in different replicas of the cell that enhance the BE, a phenomenon that does not happen for C$_2$S$^{\bullet\bullet}$ and C$_4$S$^{\bullet\bullet}$. In fact, we would expect C$_4$S$^{\bullet\bullet}$ to show larger BEs, but its elongated structure makes it challenging to maximize its contact area with the ice surface, hence interacting mostly through the C-end and pointing the S-end upwards. 

Finally, for molecules of Group IV, even though S$_2^{\bullet\bullet}$ is heavier and larger than S$^{\bullet\bullet}$, its interaction with the ice is the weakest of the entire set of the tested molecules. S$_2^{\bullet\bullet}$ has an electronic triplet ground state (like O$_2$) and is the only homonuclear molecule of the set. Its dipole moment is zero, but not its quadrupole moment, which allows the onset of a quadrupolar interaction between the S-S bond region and a dH of the ice surface. However, its BE is lower (2393 K) than that for S$^{\bullet\bullet}$ (3921 K).

\begin{table*}[htp]
    \centering
            \caption{Computed binding energies (\textit{BE disp}, in Kelvin, and without ZPE corrections) of the S-bearing species set at the P-ice (010) (2x1) super cell. The first column refers to the adsorbed species. Columns 2-4 contain the values computed at DFT//DFT level of theory. The last three columns report the same values computed at DFT//HF-3c level of theory. For the closed-shell species (computed at B3LYP-D3(BJ) level), the BE contributions arising from the non-dispersive (\textit{no disp}) and the dispersive (\textit{disp}) forces, together with its percentage (\%), is listed. For the open-shell species (computed at M06-2X level), the \textit{BE no disp} and \textit{disp} contributions cannot be separated. Whenever more than one adsorption geometry was found, all the \textit{BEs} are listed and separated by a slash.}
    \begin{tabular}{l c c c c c c }
    \hline
        \toprule
    \multicolumn{7}{c}{\textit{BEs on the crystalline water ice}} \\
    \toprule
        & \multicolumn{3}{c}{DFT//DFT} & \multicolumn{3}{c}{DFT//HF-3c} \\
    \cmidrule(r){2-4}\cmidrule(l){5-7}
        Species & \textit{BE disp} & \textit{BE no disp} & \textit{disp(\%)} & \textit{BE disp} & \textit{BE no disp} & \textit{disp(\%)}\\
    \hline
CS       &	4582        &  1600     & 2982(65\%)      &  3861         & 1239    & 2622(68\%)     \\
C$_3$S 	  & 6038 / 6723	&  1564/2129     &  4474(74\%) / 4594(68\%)     &   5713 / 6146 & 421/1443    &  5292(93\%) / 4703(77\%)    \\
CH$_3$SH    & 6182        &  2959     & 3223(52\%)      &   5340        & 2081    & 3259(61\%)    \\
H$_2$CS	  & 4847 / 6242 &  2225 / 2850     &    2622(54\%) / 3392(54\%)   &  4258 / 5641  & 1792/2417     & 2466(58\%) / 3223(57\%)      \\
H$_2$S	  & 5268        &  3332     & 1936(37\%)      &   4679        & 2862    & 1816(39\%)     \\
H$_2$S$_2$     &	5436 / 6158 &  1624/1936     & 3813(70\%) / 4222(69\%)      &  5027         &  1527   & 3500(70\%)      \\
OCS	  & 3007 / 3440 &  -229 / 325     & 3235(107\%) / 3115(90\%)      &   2790        & 132     & 2658 (95\%)      \\
SO$_2$      &	6880        &	3608    & 3271(48\%)      &    5929       & 2502    & 3428(58\%)     \\

HCS$^{\bullet}$    & 4113               &       &      &  3849           &     &      \\
HS$^{\bullet}$	 & 4270	              &       &      &  3247           &     &       \\
HS$_2^{\bullet}$   & 4270 / 4763 / 4871 &       &      &  3572 / 3139    &     &      \\
NS$^{\bullet}$	 & 4330               &	      &      &  3765           &     &      \\

C$_2$S$^{\bullet\bullet}$ & 4089 / 4691     &       &      &  3993 / 4498    &     &      \\
C$_4$S$^{\bullet\bullet}$ & 	4366 / 4414 &       &      &  4246           &     &      \\
S$^{\bullet\bullet}$   & 3921	        &       &      &  3247           &     &       \\
S$_2^{\bullet\bullet}$ & 	2393     	&       &      &  2009           &     &      \\
SO$^{\bullet\bullet}$  & 	3861        &       &      &  1924           &     &       \\
     \toprule
    \end{tabular}
    \label{tab:BE_cry}
\end{table*}

\subsubsection{ONIOM2-corrected BEs} \label{subsec:oniom}

For some species, BEs computed at DFT//DFT have been refined at a CCSD(T) level by applying the ONIOM2 methodology by \cite{dapprich:1999}, considering that most of the BE contribution arises from the local interactions between the adsorbed species and a small number of nearby water molecules. These calculations were performed on the following seven S-bearing species: CH$_3$SH, CS, H$_2$CS, OCS, SO$_2$, NS$^{\bullet}$, and S$_2^{\bullet\bullet}$. These molecules have been chosen because they well represent both closed- and open-shell species and sample the entire range of BEs of the 17 species of this work.

\begin{figure}[htb]
    \centering
    \gridline{\fig{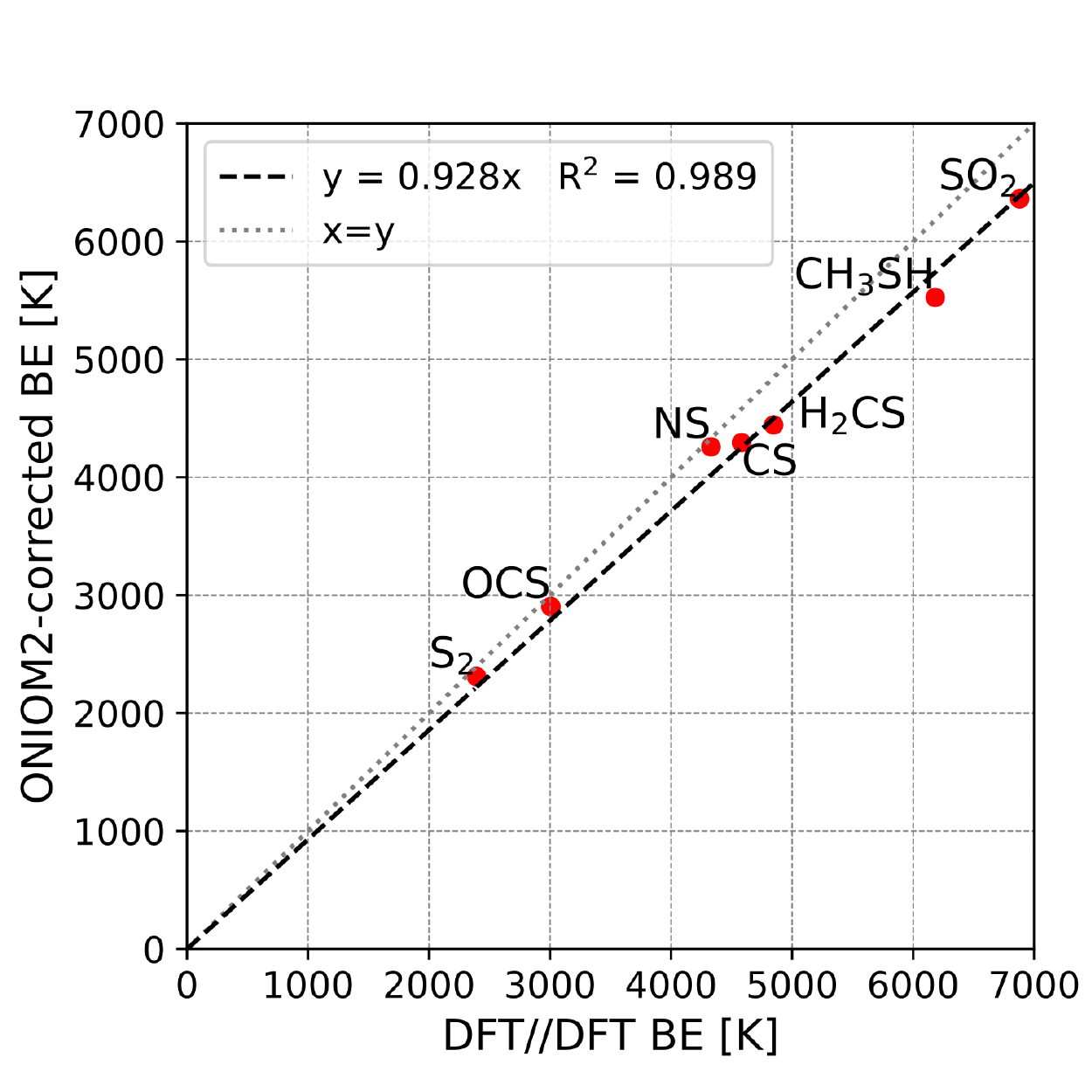}{\linewidth}{(a)}}
    \gridline{\fig{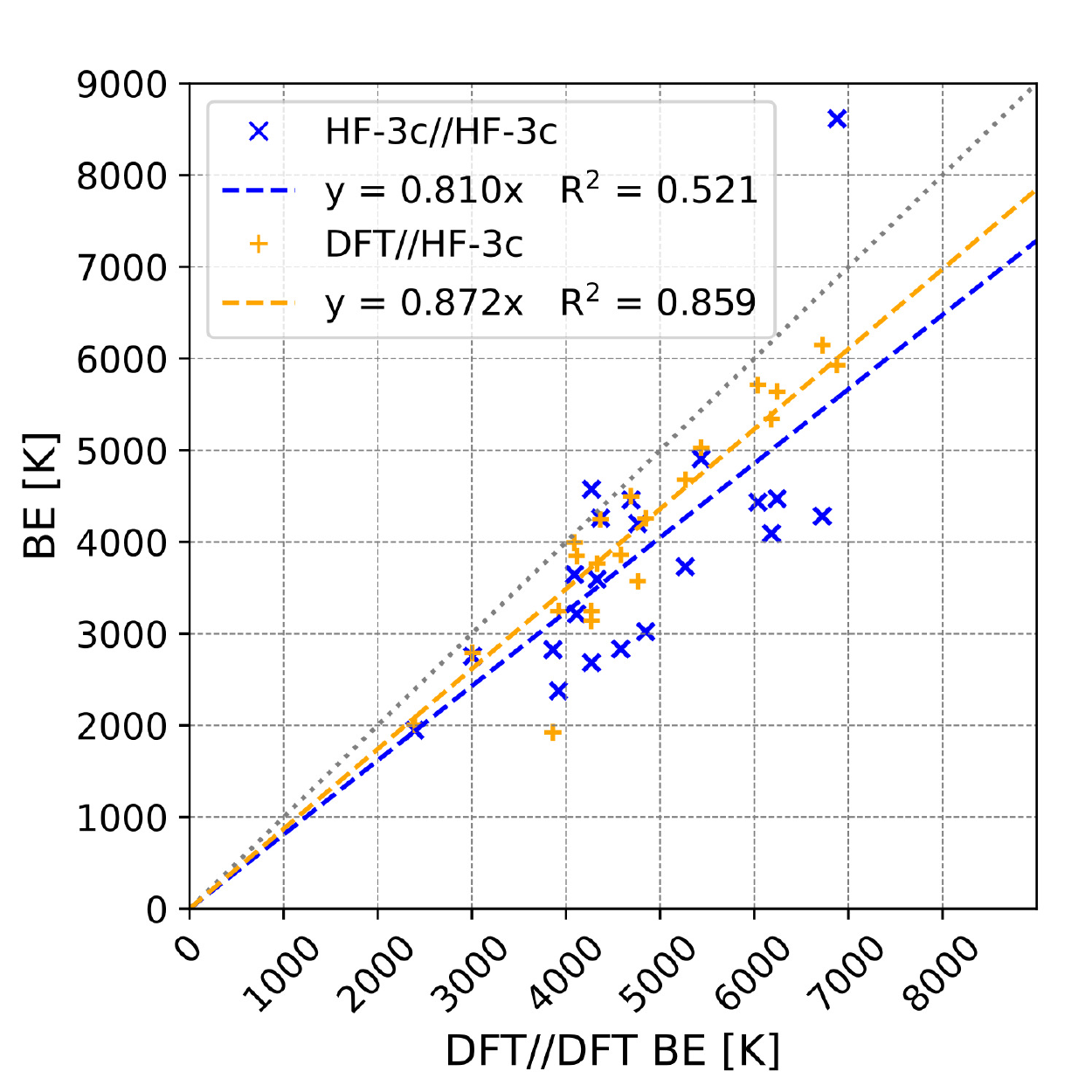}{\linewidth}{(b)}}
    \caption{(a) Best linear fit between the DFT//DFT and the  ONIOM2-corrected BEs for the crystalline ice systems. (b) Best linear fit between the DFT//DFT BEs and the  HF-3c//HF-3c (in blue) and the DFT//HF-3c (in orange) schemes, for the crystalline ice systems. BEs  in Kelvins.}
      \label{fig:hf3c_ccsdt}
  \end{figure}

Figure \ref{fig:hf3c_ccsdt} (panel a), plots the ONIOM2-corrected BEs against the DFT//DFT ones, showing a very good fit that confirms the accuracy and reliability of the DFT//DFT scheme adopted so far, with the latter slightly overestimated compared to the ONIOM2-corrected one.

\subsubsection{DFT//HF-3c BEs} \label{subsec:hf3c}

To cope with the larger size of the amorphous ice model, we tested the performance of the cheaper DFT//HF-3c scheme on the crystalline ice model to assess its applicability also for the amorphous ice.

Starting from the previously optimized DFT structures, we re-optimized the geometries at HF-3c level. In general, the HF-3c optimized geometries are comparable with the DFT ones, with the exception of very few cases, such as the SO$_2$ adsorption complex (see below). However, by comparing the HF-3c//HF-3c BEs with the DFT//DFT ones, the resulting correlation is rather coarse (see Figure \ref{fig:hf3c_ccsdt}, panel b, blue symbols and line).

However, a much better agreement is shown by the DFT//HF-3c BEs (the linear correlation improves notably, see Figure \ref{fig:hf3c_ccsdt}, panel b,  orange symbols and line). Such an agreement was already shown in \cite{ferrero:2020}, and the accuracy of the DFT//HF-3c scheme was also exhibited in molecular crystals \citep{cutini:2016}, polypeptides \citep{cutini:2017} and pure-silica zeolites \citep{cutini:2019}. 
These results, thus, extend the application of the DFT//HF-3c scheme to derive accurate BEs for S-bearing molecules adsorbed on ice surfaces.

HF-3c//HF-3c data (blue symbols) of Figure \ref{fig:hf3c_ccsdt} reveal an outlier (for SO$_2$, around 7000 K), the HF-3c//HF-3c BE being 25\% higher than the DFT//DFT one, while the DFT//HF-3c value is about 15\% smaller than the DFT//DFT value. 

This effect is because HF-3c optimized geometries show shorter distances between the adsorbate and the ice model, which, when computing the single point energy at DFT level, will increase both the dispersive interactions and the exchange repulsion contributions. For instance, HF-3c reduces the distance between the S atom of SO$_2$ and an O atom of the ice surface to 2.25 \r{A}, compared to the 2.48 \r{A} of the B3LYP-D3(BJ) optimization. Moreover, by looking at the \textit{BE disp} values of Table \ref{tab:BE_cry}, we notice that the percentage of dispersion in DFT//HF-3c BE values is almost always larger than the one of DFT//DFT BEs.

Finally, HF-3c can be critical for radical species like NS$^{\bullet}$. The NS$^{\bullet}$ experimental bond length value  is 1.494 \r{A} \citep{Peebles:2005}, while the calculated ones are 1.480 \r{A} with M06-2X and 1.651 \r{A} with HF-3c. The charge and spin density distributions are also different, with  76\% and 59\% spin density on nitrogen at HF-3c and at M06-2X, respectively. The higher spin localization of HF-3c causes the lengthening of the bond  which, when adsorbed, shrinks to 1.633 \r{A}, resulting in a gain in energy of about 10 kJ mol$^{-1}$ (at DFT//HF-3c). As this behaviour is systematic (for both free and adsorbed molecule) it cancels out when computing the BE.
In conclusion, HF-3c can be used as a cheap level of theory provided that a careful check is exerted for species with uncommon structures and electronic configurations (especially open-shell ones).

\subsection{BEs on Amorphous Ice} \label{subsec:be_amor}
The amorphous ice shows a richer variety of adsorption sites causing a distribution of BE values for the considered molecules.  We characterized eight surface binding sites (dH1-6 as H-bond donors and dO7-8 as H-bond acceptors), which were chosen as a function of their H-bonding ability. We manually built up the starting geometry, exposing the ASW electrophilic regions (due to dangling OH bonds, while their counterparts correspond to exposed O atoms) to the nucleophilic ones of the adsorbates and vice versa (following the principle of electrostatic complementarity). The distance between the adsorbate and the surface was set up as the sum of the van der Waals radii of the corresponding closest atoms, by means of computer graphics manipulation. By proceeding this way, we obtained a total of 136 adsorption complexes. As explained in Sec. \ref{subsec:method}, the complexes were first optimized at the HF-3c level followed by a harmonic frequency calculation to confirm all structures as minima.
Table \ref{tab:BE_am} shows the BEs value at the final DFT//HF-3c level. The mean ($<$\textit{BE}$>$) and standard deviation ($\sigma$) for the BEs are also provided, although we recommend the reader not to consider them as representative of the ensemble of BEs here listed, given the limited sampling of binding sites performed in this study, far from being suitable for a statistical treatment.

From the analysis of our data, it stands out that dH4 and dH6 sites, residing in the cavity domain of the ice model, exhibit  the highest BE values due to the close proximity of the surrounding water molecules. The reverse happens for dO7 and dO8, placed on the bottom and as outermost water molecules of the model, which are the least favourable binding sites. 

On average, the dispersion contributions to the BEs are higher when the adsorption take place at the amorphous ice than at the crystalline ice due to closer contacts of the adsorbate with the ice surface. 
While in the crystalline proton ordered P-ice model the H-bond chains extend to infinity ensuring a reinforcement of the H-bonding strength (cooperativity effect), this is not the case for the ASW model. In ASW, the random organization of the water molecules breaks the H-bonding chain extensions, reducing the H-bonding cooperativity and, therefore, its strength. This, in turn, also affects the H-bonding strength exhibited by the chain terminal ice OH groups involved in the interactions with the adsorbates, decreasing the corresponding BE values. This is shown clearly by the higher BE values computed for adsorption on the crystalline P-ice with respect to the ASW model, in agreement with previous literature data \citep{ferrero:2020}. However, other effects, like the easier structure deformation for the amorphous ice, allows accommodating the adsorbates closer to the ice surface maximizing both the H-bond strength and the dispersive contribution, overcoming the loss of H-bonding cooperativity. 

The above interpretation  can explain the presence of some very high BE values, which depart from the rest of the BEs of the corresponding ensembles. This happens for CH$_3$SH, C$_3$S,  H$_2$CS, H$_2$S, H$_2$S$_2$, C$_2$S$^{\bullet\bullet}$, and C$_4$S$^{\bullet\bullet}$, adsorbed in the dH2 and dH5 sites (located at the edges of the top surface). In these cases a dramatic structural rearrangement of the amorphous ice took place, with a global energy gain that can reach 2400 K, contributing to the ``anomalous'' high BEs. In practice, the species/ice interactions cause a large restructuring of  the amorphous ice, whose new structure has a lower energy than the initial ice taken as a reference.

This phenomenon is not usual when dealing with rigid structures (e.g., metal surfaces, oxides, graphene); however, soft matter like water ice is held together by the same kind of interactions keeping the adsorbate attached at the surface. Therefore, the geometrical relaxation induced by the adsorption can rearrange the structure in such a way that, once optimized, it falls in a new minimum with a lower energy with respect to the reference system. A full analysis of this phenomenon has been recently described in \cite{tinacci:2022}.

To cope with this ``artificial'' increment of the binding energy we adopted the new ice structure as a reference for the calculation of the BE, which are now coherent with the general BE ensembles.
We think this is an important issue, not common in the adsorption phenomena on metallic, ionic, or covalent surfaces, which are dominated by stronger forces than those occurring with soft matter like water ices. This problem would be mitigated only by enlarging the ice model at an increasing computational cost.

\begin{table*}[htp]
    \caption{Summary of the DFT//HF-3c \textit{BEs} (in Kelvin, and without ZPE corrections) of the 17 S-bearing species  at the amorphous ice periodic model. For each species, 8 BEs value, alongside their mean ($<$\textit{BE}$>$) and standard deviation ($\sigma$)} are provided. 
    \centering
    \begin{tabular}{l@{\hskip 0.3in} c@{\hskip 0.3in} c@{\hskip 0.3in} c@{\hskip 0.3in} c@{\hskip 0.3in} c@{\hskip 0.3in} c@{\hskip 0.3in} c@{\hskip 0.3in} c@{\hskip 0.3in} c@{\hskip 0.3in} c}
    \toprule
    \multicolumn{11}{c}{\textit{Amorphous Ice BE Values}} \\
    \hline
    Species & dH1 & dH2 & dH3 & dH4 & dH5 & dH6 & dO7 & dO8 & \textit{$<BE>$} & \textit{$\sigma$} \\
    \hline
CS	& 4270   &	1852   &	3596  &	3283  & 	3380 &	3584  &	1155  &	1648 & 2846	& 1054  \\
C$_3$S	& 5472   &	6014   &	6122  &	5412  & 	5112 &	5941  &	1275  &	2959 & 4788	& 1630  \\
CH$_3$SH	& 3608   &	4991   &	3584  &	3271  & 	3271 &	4630  &	2622  &	2850 & 3603 &	769 \\
H$_2$CS	& 4161   &	4967   &	3416  &	3488  & 	4378 &	4053  &	1960  &	3115  & 3692	& 859 \\
H$_2$S	& 3452   &	5208   &	3283  &	4210  & 	4630 &	2622  &	2285  &	2309  & 3500	& 1026\\
H$_2$S$_2$	& 3007   &	4234   &	4558  &	4462  & 	3801 &	3801  &	3175  &	4883 & 3990	& 623 \\	
OCS	& 2875   &	1491   &	2574  &	2177  & 	1660 &	3320  &	2057  &	3175 & 2416	& 637     \\
SO$_2$	& 3344   &	3175   &	3644  &	6759  & 	3680 &	4426  &	2730  &	2442  & 3775	& 1263 \\

HCS$^{\bullet}$	& 2117   &	2225   &	2454  &	2117  & 	1491 &	3560  &	2478  &	2201 & 2330 &	545   \\  
HS$^{\bullet}$	& 3163   &	2454   &	2057  &	2646  & 	5003 &	2177  &	1624  &	1251    & 2547	& 1080  \\
HS$_2^{\bullet}$	& 2562   &	1527   &	2586  &	3031  & 	2670 &	2550  &	2778  &	2694               & 2550 &	413    \\
NS$^{\bullet}$	& 3283   &	1431   &	1828  &	1768  & 	3416 &	3428  &	1768  &	1876  & 2350	& 805   \\	

C$_2$S$^{\bullet\bullet}$	& 4426   &	3163   &	5088  &	3572  & 	2009 &	4991  &	2009  &	2393 & 3456	& 1193\\ 
C$_4$S$^{\bullet\bullet}$	& 5075   &	2165   &	5797  &	3716  & 	4943 &	5569  &	4306  &	4077 & 4456	& 1097  \\		
S$^{\bullet\bullet}$ 	& 3247   &	2081   &	2550  &	2321  & 	2526 &	2321  &	1852  &	1852 & 2344	& 425  \\	
S$_2^{\bullet\bullet}$	& 1924   &	1455   &	2850  &	1552  & 	2009 &	2069  &	1648  &	1203 & 1839 & 472 \\ 
SO$^{\bullet\bullet}$	& 4041   &	1082   &	2634  &	1419  & 	1696 &	2345  &	1672  &	1696 & 2073	& 874  \\ 
  	
\toprule
    \end{tabular}
    \label{tab:BE_am}
\end{table*}

\subsection{Zero-Point Energy (ZPE) corrections} \label{subsec:zpe}

\begin{figure}[htb]
    \centering
    \includegraphics[width=\linewidth]{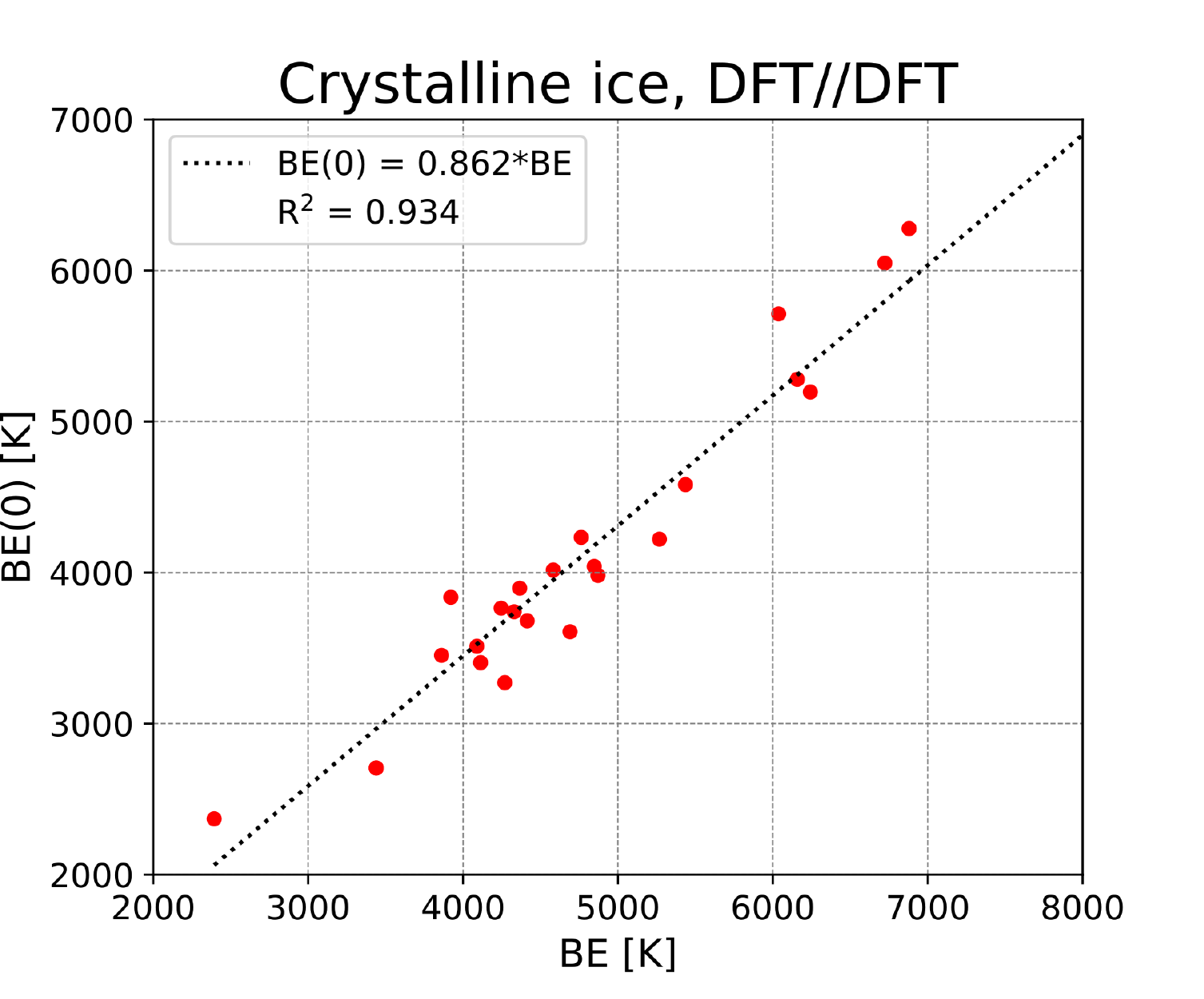}
  \caption{Analysis of ZPE corrections. Correlation between DFT//DFT BE(0) and BE for the adsorption at the crystalline ice. The intercept is set to zero.}
  \label{fig:zpe}
  \end{figure}

As anticipated in Section \ref{subsec:method}, the ZPE corrections were computed following two different schemes, depending on the adopted surface models. 
For the adsorption on the crystalline ice, we computed the harmonic frequencies of a fragment consisting of the S-bearing species and the two closest water molecules of the ice with the assumption that deformation occurs only in the proximity of the adsorption site. 
The $\Delta$ZPE is computed following the advice described in Appendix, \S ~\ref{subsec:be} to arrive at ZPE-corrected BE, i.e., BE(0). Figure \ref{fig:zpe} shows a good linear correlation between BE and BE(0), showing the average ZPE correction being about 14\% of the BE.

The final BE(0) values for all the adsorption complexes at the amorphous ice have, therefore, been computed by multiplying the DFT//HF-3c BEs by a scale factor of 0.862 (see Figure \ref{fig:zpe}, panel (a)).

\section{Discussion} \label{sec:discussion}

\subsection{Comparison with BEs for non S-bearing adsorbates} \label{subsec:ferrero}

In the recent work by \cite{ferrero:2020}, the adsorption of 17 closed-shell and 4 open-shell species was simulated. The list included: i) non-dipolar molecules with significative quadrupole moments (i.e., H$_2$, N$_2$, and O$_2$) or higher-order multiple moments (i.e., CH$_4$), hence interacting weakly with the ice; ii) molecules that, besides exhibiting a quadrupole moment, established H-bonds with surface dH atoms (i.e., CO, OCS, and CO$_2$); iii) amphiprotic molecules, that are both H-bond acceptors and donors (i.e., NH$_3$, H$_2$O, HCl, HCN, and H$_2$S); iv) larger molecules that are capable of establishing a greater number of interactions with the surface due to the presence of several functional groups, hence giving rise to the highest BE values of the set (i.e., CH$_3$OH, CH$_3$CN, H$_2$CO, HCONH$_2$, and HCOOH); and v) the open-shell species OH$^{\bullet}$, NH$_2^{\bullet}$, CH$_3^{\bullet}$, and HCO$^{\bullet}$, the two former ones forming strong H-bonds with dH and dO sites, at variance with the two latter ones. 

\begin{figure}[htb]
    \centering
    \gridline{\fig{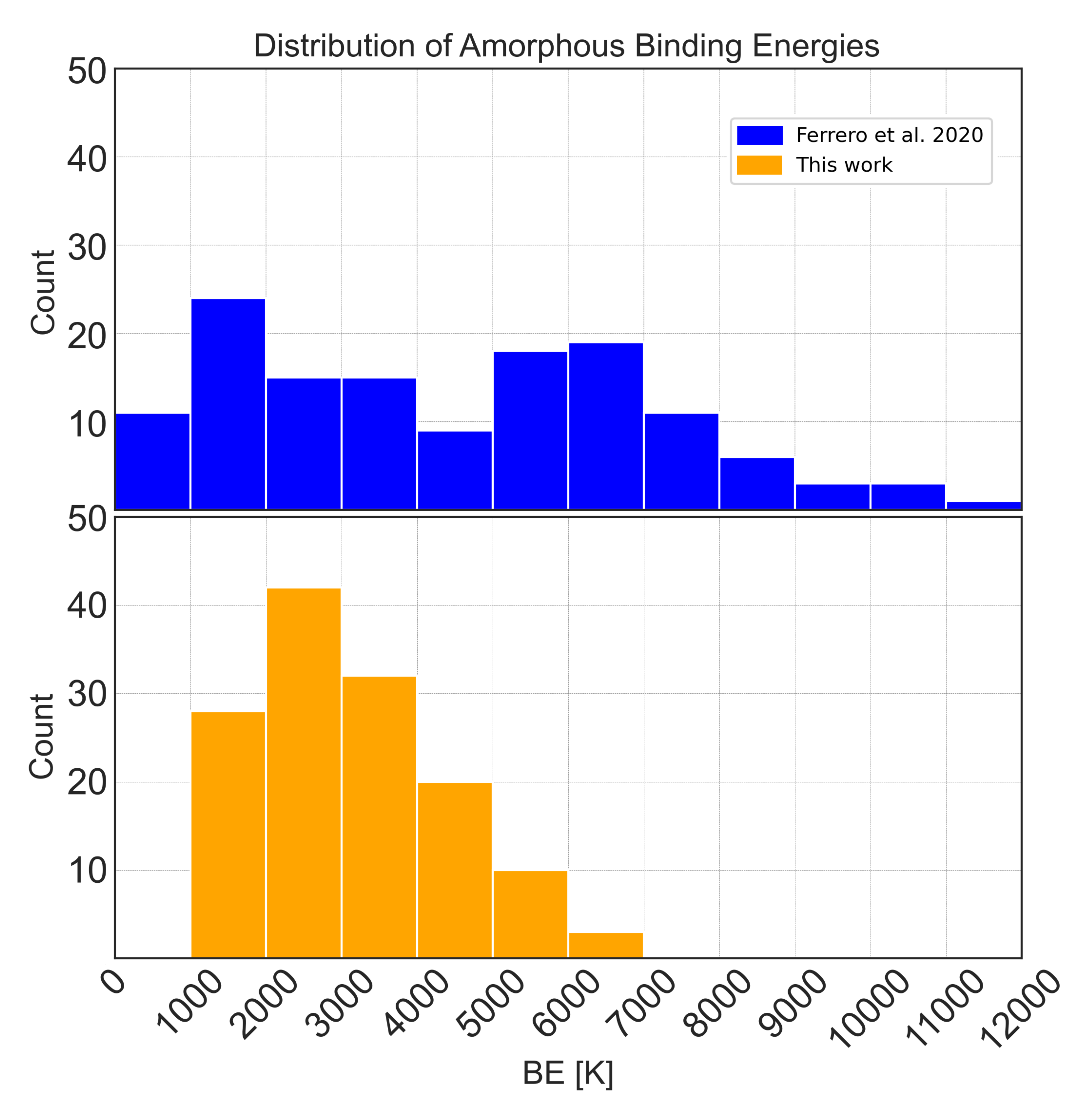}{\linewidth}{(a)}}
    \gridline{\fig{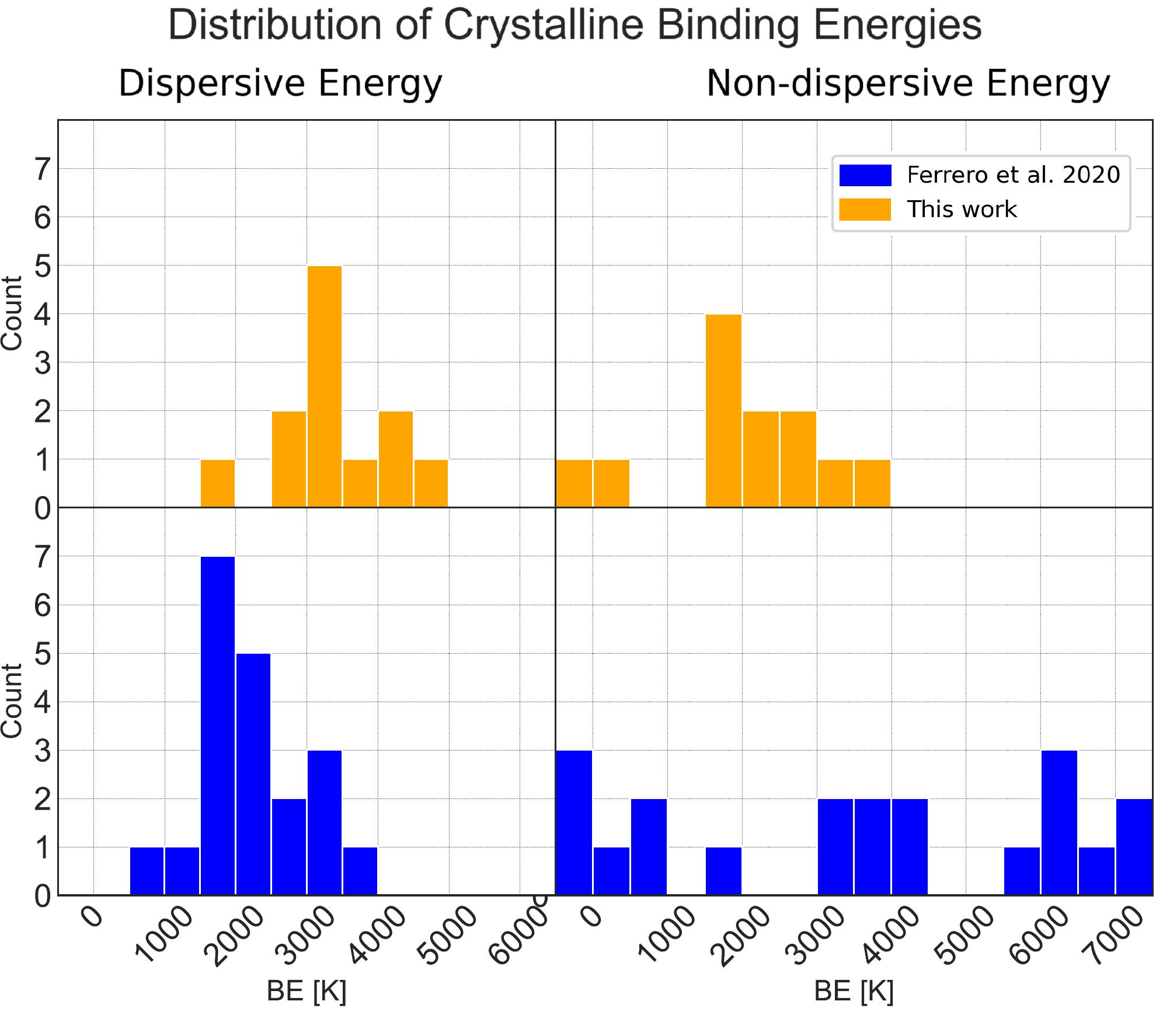}{0.9\linewidth}{(b)}}
  \caption{(a) Distribution of BEs computed on the amorphous ice model for S-bearing species (orange) and the species of \cite{ferrero:2020} (blue). (b) Distribution of the dispersive (left panel) and non-dispersive (right panel) contributions to the crystalline BEs for closed-shell species. }
  \label{fig:distribution}
  \end{figure}

Figure \ref{fig:distribution} compares the BEs of the S-bearing species with those of \cite{ferrero:2020}. The distribution of BEs in \cite{ferrero:2020} has a bi-modal character (see panel a), indicating two different adsorption behaviors in the set of the considered species: a group of low BEs (characterized by dispersive interactions), and a group of high BEs (amphiprotic and larger species involved in H-bond). In contrast, S-containing species span a much narrow range of BEs, suggesting a common adsorption feature, despite the diverse chemical structures.

Deeper insights into the contributions to the BE of these two sets of species arise from the analysis of the closed-shell species adsorbed at the crystalline ice model. Each BE can be decomposed in dispersive and non-dispersive contributions (see panel b of Figure \ref{fig:distribution}) for all considered species.

The distribution of the dispersive contributions highlights two main facts: i) this term is more important when heavy atoms (like sulphur) belong to the adsorbed species;  ii) the mean value of the distribution for S-bearing species (ca. 3500 K) is higher than that of the non-dispersive terms (ca. 2000 K), as dispersion accounts for more than 50\% of the BEs provided in this work. Therefore, weak interactions are the most relevant and crucial forces in the description of the adsorption of S-containing species on ices and, thus, they need to be carefully accounted for to properly describe the sulphur chemistry on icy mantles.  The above considerations are not surprising due to higher atomic polarizability of second row atoms compared to C-, N- and O-containing molecules.

The different behavior of the two sets of species will have consequences on their chemistry. Considering the BE ensembles, one can argue S-containing species to show higher desorption temperatures compared to weakly interacting molecules, like O$_2$, CH$_4$ or CO. On the contrary, S-bearing species interact weaker than amphiprotic molecules with ice. The same inference applies to their surface mobility. As usually, the barriers for diffusion are taken as an arbitrary fraction of the BEs \citep{cuppen:2017,kouchi:2020}, we can assume that species with lower BEs will diffuse faster than those with higher BEs, setting up constrains on the reactions that can take place on an icy surface. The particularly low BEs of S$_2^{\bullet\bullet}$ (the weakest bounded S-bearing species) also implies its highest mobility. In the hypothesis that polysulphanes could be a reservoir of sulphur, the high mobility of S$_2^{\bullet\bullet}$ could favour their encountering and reaction, forming chains and cyclic sulphur species \citep{druard:2012,mifsud:2021} that hardly desorb due to the very large dispersive contribution to their BE.

\subsection{S-bearing vs O-bearing species}
\begin{table*}[htp]
    \caption{Contribution to selected BE(0) (in Kelvin) for S-bearing species versus O-bearing species adsorption at crystalline ice. The latter are taken from \cite{ferrero:2020}. The BE(0) are decomposed into \textit{no disp} and \textit{disp} contributions, applying the percentage of dispersion (\%) obtained for each species. The dipole moment $\mu$ (in Debye) of the free molecules is computed at B3LYP-D3(BJ)/A-VTZ* level of theory. When more than one BE was present, the average value is reported.}
    \begin{tabular}{l c c c c | l c c c c}
    \toprule
    \multicolumn{5}{c}{Sulphur bearing species} & \multicolumn{5}{c}{Oxygen bearing species} \\
    \hline
    \hline
    Species & \textit{BE(0)} & \textit{no disp} & \textit{disp(\%)} & $\mu$ & Species & \textit{BE(0)} & \textit{no disp} & \textit{disp(\%)} & $\mu$ \\
CS & 2453 & 785 & 1668(68\%) & 1.9 & CO & 1663 &  50 & 1613(97\%) & 0.1 \\
CH$_3$SH	&  4603  & 1795  &	2808(61\%) & 1.5 &	CH$_3$OH  & 7385 & 5391 & 1994(27\%)  & 1.7 \\
H$_2$CS	&  4267  & 1792  &	2475(58\%) & 1.7 &	H$_2$CO  & 5187 & 3268 & 1919(37\%) & 4.6 \\
H$_2$S	&  4033  & 2460  &	1573(39\%) & 1.1 &	H$_2$O  & 7200   & 5832 & 1368(19\%) & 2.0 \\
OCS	&  2405  &  120 &	2285(95\%) & 0.8 &	CO$_2$  & 2568  & 796 & 1772(69\%)& 0.0\\
\toprule
    \end{tabular}
    \label{tab:S_vs_O}
\end{table*}

Besides these general considerations, a punctual comparison between S-bearing and O-bearing species can be drawn for some cases in order to highlight a few exceptions to what we generally understood about our set of molecules. 

In Table \ref{tab:S_vs_O} we compare the contributions to the BE(0) values of some S-bearing species with their oxygen analogues. As a rule of thumb, we would expect molecules containing sulphur to always show a BE(0) lower than the corresponding O-bearing ones, with the dispersive interactions being larger for the former group. However, by considering case by case, we notice that there are some exceptions to this behaviour. For the pairs CS/CO the trend is the opposite, while the OCS/CO$_2$ pair shows many similarities. 

Indeed, CS exhibits a larger BE(0) and a lower percentage of dispersive energy than its O-analogue, CO. This can be explained by looking at the dipole moments of these species. CS shows a large dipole moment (1.9 D), at variance with the negligible one of CO. The H-bonding interaction of the two species with a dangling OH is therefore stronger for CS, as it is the dispersive interaction of the S atom with the water molecule of the surface (see figure in Appendix, \S ~\ref{subsec:geom}).

Both OCS and CO$_2$ have a linear geometry, an almost null dipole and interact through an H-bond with a dangling OH of the surface. While OCS exhibits a larger dispersion interaction, for CO$_2$ the electrostatic component makes up for about 30\% of the binding energy, yielding a comparable BE(0) for the two molecules. 

On the other hand, the other pairs behave as we expected and showed in the previous section \ref{subsec:ferrero}. The adsorption geometries show some similarities between the two members of each pair, with the most important differences due to the presence and strength of H-bonds. Specifically,  CH$_3$SH/CH$_3$OH and H$_2$S/H$_2$O exhibit the same number of H-bonds, while for H$_2$CS/H$_2$CO only the latter establish a H-bond with the surface through the C=O group.

\subsection{Comparison with literature} \label{subsec:literature}

\begin{table*}[htp]
\caption{Summary of the computed BE(0) values (in Kelvin) in comparison with data from the literature. The second column contains the BEs on the crystalline P-ice (010) model, the third and the fourth column contain the minimum and maximum BE value found for the amorphous ice, the fifth and the sixth list the mean ($<\textit{BE}>$) and the standard deviation ($\sigma$) relative to the amorphous BEs,  while columns from 7 to 11 report BEs from the literature, respectively from computed data  (\cite{das:2018} and \cite{wakelam:2017}), from databases (KIDA \citep{wakelam:2015} and UMIST \citep{mcelroy:2013}) and from experiments (\cite{penteado:2017}). The BE(0) values were obtained by applying the equation 0.862*BE=BE(0) (see text for more details).}
    \centering
    \begin{tabular}{l c c c c c c c c c c}
    \hline
     \multicolumn{11}{c}{Binding Energies} \\
    \hline
    & \multicolumn{5}{c}{This work} & \multicolumn{5}{c}{Literature} \\
    \cmidrule(r){2-6}\cmidrule(l){7-11}
& Crystalline & \multicolumn{4}{c}{Amorphous} & \multicolumn{2}{c}{Computed} & \multicolumn{2}{c}{Database} & Various \\
Specie	& \textit{BE(0)} & \textit{Min} & \textit{Max} & \textit{\textit{$<BE>$}} & \textit{$\sigma$}	& \textit{Das}  & \textit{Wakelam}  & \textit{KIDA}  & \textit{UMIST} &	\textit{Penteado} \\		
\hline
CS  &	 2453  & 995    &    3680 & 2453 &	909 &
 	2217 & 3200  &	3200  & 1900 &	1800 ± 500 \\  
C$_3$S 	  &  4925 / 5298   & 1099   & 5277 & 4128 &	1405 
& &  &	3500  & 3500 &	3000 ± 500 \\
CH$_3$SH &   4603  &  2260   &	4302  & 3106 & 663
& & 4200 &  &  & \\
H$_2$CS	  & 3670 / 4863    & 1690   &	4282 & 3183 &	740
&	3110 & 4400  &	4400 & 2700 &	2025 ± 500 \\
H$_2$S	  &  4033   & 1970   &	4489  & 3017 & 884
&	2556 & 2900  &	2700  &	2743 & 2290 ± 90 \\
H$_2$S$_2$  & 4333  &	2592   &	4209 & 3439 & 537
&	4368 &   &	3100 & 3100 &	2600 ± 500 \\
OCS	  &  2405   & 1286    &	2861   & 2083 & 549
&	1571 &  2100  &	2400 & 2888 &	2325 ± 95 \\
SO$_2$  &	 5111   & 2105   &	5827 & 3254 & 1089
&    	3745  & 5000 &	3400  &	5330 & 3010 ± 110 \\

HCS$^{\bullet}$	 &   3318  & 1286     &	3069  & 2009 &	469
&	2713 & 2900  &	2900 & 2350 &	2000 ± 500 \\  

HS$^{\bullet}$	 &  2799   & 1078    &	4313  & 2195 &	931
& 2221 & 2700 &	2700 & 1500 &	1350 ± 500 \\  
HS$_2^{\bullet}$ &   2706 / 3079   &  1317  &	2613 & 2198 & 356
&	4014 &  &	2650 & 2650 &	2300 ± 500 \\
NS$^{\bullet}$	 &  3245  &   1234 &	2955   & 2026 & 694
&	2774 &  &	1900 &	1900 & 1800 ± 500 \\

C$_2$S$^{\bullet\bullet}$	 & 3442 / 3877    & 1731    &	4385  & 2979 &	1028
&	2447 & &	2700 & &	2500 ± 500 \\
C$_4$S$^{\bullet\bullet}$ & 3660	    & 1866  & 4997 &	 3841 &	945
& &	&	4300 &	4300 & 3500 ± 500 \\
S$^{\bullet\bullet}$  &   2799  &  1579    &	2799  & 2020	&366
& 1428 & 2600 &	2600 & 1100 &	985 ± 495 \\
S$_2^{\bullet\bullet}$ & 1732	    &  1037  &	2457     &	1644 & 1585	& 407
&  &	2200 & 2200 &	2000 ± 500 \\
SO$^{\bullet\bullet}$ & 	1658    &  933   &	3483     &	2128 & 1787 &	754
& 2900 &	2800 &	2600 & 1800 ± 500 \\

\hline
    \end{tabular}
    \label{tab:BE_literature}
\end{table*}

Before comparing the present results with literature experimental data, some general considerations are at hands. In the ISM, water ice mantle is the result of the \textit{in situ} water formation through surface reactions. Therefore, the final ice structure is the result
of the energetic locally hot spots injected in the mantle where intermediates (e.g., OH$^{\bullet}$, HOO$^{\bullet}$, etc.) and final products are formed. As the injected energy is large, it locally heats up the ice, shaping the final structure of the grain in a complex way, difficult to attain through laboratory experiments. In the terrestrial laboratories, ice is prepared directly by deposition from vapor water molecules, with a certain kinetics, on cold metal fingers \citep{penteado:2017}. The final ice can be amorphous (either compact or porous) or crystalline, depending on the kinetics of adsorption and processing temperature. Unfortunately, the detailed structure of amorphous ice can only be partly inferred indirectly by spectroscopic measurements, therefore limiting our knowledge of the atomistic details. The same difficulty occurs for computer modeling: indeed, we are not aware of computer simulations reacting hydrogen and oxygen atoms step by step up to form a sizable icy grain, even if a simpler attempt to form a single water molecule on the forsterite surface has been recently reported \citep{molpeceres_2019}.

As a function of the adopted ice models in computer simulation, the external surface may show a variety of different kind of adsorption sites, giving rise to  different BEs \citep{song:2017}. When very few sites are available, like in the proton ordered ice model adopted here, no more than two different BE values resulted. For the amorphous model, instead, a significant variability of binding sites resulted, giving rise to an ensemble of BEs.  
As reported in details by \cite{ferrero:2020} and not repeated here, extra care should be exerted when comparing the computed BEs with the ones from experiment.

For instance, while BEs are straightforwardly calculated by computer simulations (see \S ~\ref{subsec:be}), experimental BEs are never directly observed in the thermal desorption, as TPD peaks are usually interpreted through the \cite{redhead:1962} method or more sophisticated techniques, through which desorption adsorption energies (DAEs) are worked out. As DAEs may  be a function of the surface coverage $\theta$, the comparison with the theoretical BEs is not straightforward (for a recent review see \cite{Minissale:2022}). For instance, the ice structure is sensitive to both temperature and adsorption/desorption processes (at variance with surfaces from covalent/ionic solids) as its layers are held together by weaker interactions (e.g., H-bonds in water ices or dispersion in CO ices), which are similar in strength to the BE values. Obviously, ice restructuring processes may affect the final value of the DAEs. 

\begin{figure}
    \centering
    \includegraphics[width=\linewidth]{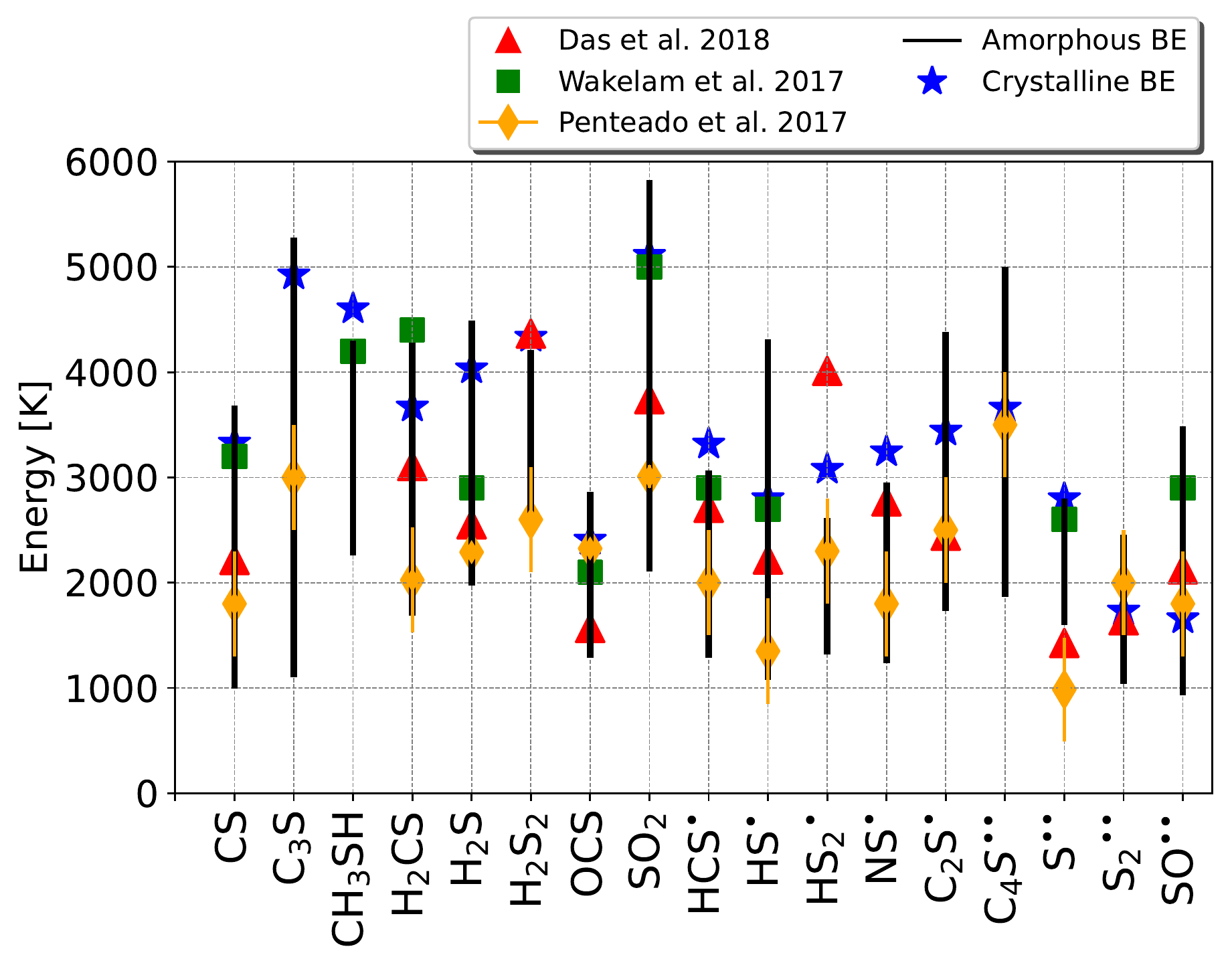}
    \caption{Computed and literature BEs (in Kelvin). For each species, the solid black line covers the min/max  ZPE-corrected BE values for the amorphous ice model, while blue stars correspond to the ZPE-corrected BEs on the crystalline ice model.}
    \label{fig:close+open}
\end{figure}

In the following, our BEs are compared with those from \cite{penteado:2017} (see Figure \ref{fig:close+open}). These authors showed a list of recommended BEs, collecting data from previous works and providing an uncertainty range for each value. For H$_2$S, OCS, and SO$_2$, the BEs are derived \textit{a posteriori}  from the experimental TPD measurements of \cite{collings:2004}, using the empirical relationshi of 
\begin{equation}
    BE(X) : T_{des}(X) = BE (H_2O) :T_{des}(H_2O)
\end{equation}
where X is the considered chemical species and T$_{des}$ is the desorption temperature of the species deposited on a water ice extrapolated from the TPD measurements of \cite{collings:2004}. The reference BE(H$_2$O) is set to 4800 K from previous works (see \cite{penteado:2017} for more details). The uncertainty assigned to the BE values of H$_2$S, OCS, and SO$_2$ was 3.5\%. For the other species, the provided data are based on the work of \cite{hasegawa:1993}, which in turn were based on previous works, including \cite{allen:1977}. In the latter, the BE of a molecule was treated as the sum of the binding energies of the constituent atoms to infer the interaction energy of several species on SiO$_2$ surfaces, which was subsequently scaled by an empirical factor to obtain BEs on water ice. The uncertainty on these values is set to half the BE for species whose BE is less than 1000 K and to 500 K for all other cases. 

The comparison between our computed BEs for the crystalline ice and the literature data shows, as expected, discrepancies. However, comparison of the range provided in \cite{penteado:2017} with our BE ensemble reveals that disagreement is only for S$^{\bullet\bullet}$. The additive principle applied by \cite{allen:1977} implies that S$_2^{\bullet\bullet}$ BE must be twice the value of the isolated S$^{\bullet\bullet}$ atom. However, we have seen that S$^{\bullet\bullet}$ and S$_2^{\bullet\bullet}$ showed a very similar distribution of BEs, due to the same nature of their interaction with the surface. For the other species, the ranges provided by \cite{penteado:2017} lay within our ensembles, especially for H$_2$S, OCS, and SO$_2$, which have a very narrow distribution of BEs. However, computed BEs for crystalline ice compare poorly with the average values provided by \cite{penteado:2017}, showing that the additive principle is not suitable for the description of the interactions between polyatomic species and extended ice models. 

In addition to the experimental data,  comparison is carried out with recent BEs from computational works  for a large number of species, some of which containing sulphur (see Figure \ref{fig:close+open}).

In \cite{wakelam:2017}, a single water molecule was adopted as ice model (M06-2X/aug-cc-pVTZ theory level, neither ZPE nor BSSE corrections). The lost contributions from a real ice surface were somehow recovered by a clever fit between computed BEs and experimental ones for a subset of species. The resulted fitted parameters were used to derive all other BE values. Comparison between our crystalline BEs with those of \cite{wakelam:2017} shows a difference of about 30\% of the value. A better agreement with \cite{wakelam:2017} values is shown by the BE range spanned by our amorphous BE ensembles. 

\cite{das:2018} computed the BEs for a molecular set including a hundred species, adopting a water tetramer and, in few cases, a water hexamer as ice model, at the MP2/aug-cc-pVDZ theory level, without BSSE and ZPE corrections. Among the S-bearing molecules, only H$_2$S and OCS were adsorbed on a water hexamer. The ice models employed in \cite{das:2018} still lack the effect of H-bond cooperativity (present in our periodic models) and could be the reason of the small values of BEs  for the S-containing species, even lower than those estimated by \cite{wakelam:2017}. 
Moreover, H$_2$S and OCS BEs show an increment when changing the model from the water tetramer to the water hexamer and it is not clear for how long this trend would continue before reaching a converged value. The \cite{das:2018} values are within the BE ensembles computed for the amorphous ice. Altogether, this indicates that the choice of the ice model and of the computational method have a strong relevance on the predicted BEs. Thus, more studies are needed to fully characterize BE distributions, by using larger models and exploring more binding sites.

\subsection{Binding energies and snow lines in protoplanetary disks}

As mentioned in the Introduction, the BEs of the species on ice mantles have a profound impact on the interstellar chemistry and, more specifically, in the chemical composition of the regions where planetary systems, like our own, are formed. 
A very important case is represented by the gaseous versus solid chemical composition in protoplanetary disks, which are the sites where planets, asteroids and comets eventually form. 
If a species is in the solid form then it will likely be incorporated in the (aforementioned) forming rocky objects, whereas, if it is in the gaseous form, it will only enrich the giant gaseous forming planets of the elements contained in the species \citep[e.g.][]{oberg:2021}.
Sulphur, on the other hand, seems to be a very important element for terrestrial life and, perhaps, even a key element for its emergence \citep[e.g.][]{shalayel:2020}.

As discussed in the Introduction, the major reservoirs of sulphur in cold molecular clouds (\S ~\ref{subsec:sulphur}, from where the evolution toward a planetary system starts), are unknown, though it is suspected to be in solid organo-sulphur compounds and polysulphates \citep{laas:2019,druard:2012}.
During the evolution, the quantity of sulphur in the gaseous phase changes by orders of magnitudes so that it is very important to know it in the protoplanetary disks. So far, there are no measurements for solid, and only very few for gaseous S-bearing species in protoplanetary disks (see below).
The locus where the transition of a species from the solid to the gas phase takes place is usually called “snow-line” in the astrophysical jargon, particularly so in the protoplanetary disks, which are objects relatively cold and dense \citep[e.g.][]{oberg:2021}.
Then, one efficient way to measure the solid versus gaseous content of S-bearing species in protoplanetary disks is to observe where the species disappear from the gas-phase, because they freeze onto the dust grain water-rich mantles, namely, to measure their snow-lines.
Two major parameters determine where the snow-line lies: the temperature of the dust, and the BE of the species.
In general, the larger the distance from the disk center (heated by the central forming star), the colder the dust. Therefore, species with lower BE remain gaseous at larger disk radii.

The ALMA-DOT survey of young disks provides the first preliminary observational constraints on the position of the snow-lines of some of the S- and O-bearing molecules considered in this study \citep{podio:2020a}. 
In particular, ALMA-DOT observed a protoplanetary disk, IRAS 04302+2247, seen edge-on, which makes it ideal to constrain the snow-lines of different species.
These observations show that the molecular emission is bright in an intermediate disk layer (the so-called molecular layer: \cite{aikawa:1999}), while it decreases in the disk mid plane due to molecule freeze-out onto the dust grains. 
More relevant to this work, CS, H$_2$CO, and H$_2$CS show a similar spatial distribution, and a sharp decrease of their emission is observed at smaller radii than CO \citep{podio:2020b,codella:2020}.
The estimated snowline is at ~ 25 au for H$_2$CO (and consequently H$_2$CS and CS) and ~100 au for CO \citep{vanthoff:2020b}.
The larger snow-line radius of CS with respect to that of CO is in general agreement with the two estimated BEs: 995–3680 K (CS, this work) versus 1109–1869 K (CO, \cite{ferrero:2020}).
On the other hand, the observations suggest similar snow-lines for CS, H$_2$CO, and H$_2$CS, while the estimated BEs of H$_2$CO and H$_2$CS are larger than that of CS: 3071–6194 K (H$_2$CO, \cite{ferrero:2020}) and 1690–4282 K (H$_2$CS, this work) versus 995–3680 K (CS, this work).
We emphasise, however, that the ALMA-DOT observations are unable to resolve the inner 25 au of the disk.
Therefore, higher spatial resolution observations are needed to constrain the snow-line of the species with large BEs, such as H$_2$CO and H$_2$CS.

In conclusion, the admittedly scarce observations so far available seem to be in agreement with the estimated BEs obtained through theoretical quantum chemical calculations as provided in \cite{ferrero:2020} and this work, which encourage us to pursue the work to a larger sample of species, both S- and N- bearing and the so-called interstellar Complex Organic Molecules (iCOMs, \cite{ceccarelli:2017}).

\section{Conclusions} \label{sec:conclusion}
In the present work, the binding energies (BEs) of 17 astro-chemically relevant S-bearing species adsorbed on periodic water ice models have been computed using density functional theory (DFT) based on the hybrid B3LYP-D3(BJ) (for closed-shell species) and hybrid meta-GGA M06-2X (for open-shell species) with an Ahlrichs-VTZ* double-zeta polarized gaussian basis sets. We adopted either a crystalline proton-ordered ice model and a more realistic amorphous ice model, both treated within the periodic boundary conditions. The adopted DFT results have also been refined towards a CCSD(T) level through the ONIOM2 approach carried out on the crystalline model, showing a very good convergence between the DFT BEs and those refined with the ONIOM2 scheme. As full DFT is too costly for the simulation of adsorption on the amorphous ice, we checked the reliability of the cheaper HF-3c method by contrasting the BEs with that at DFT level for the crystalline ice. While HF-3c BEs poorly correlate with the DFT ones, the BEs at single point DFT//HF-3c level give excellent correlation with full DFT values. The validated DFT//HF-3c method was used to study the adsorption process on the amorphous ice model, in which the richness of adsorption sites (up to 8) implies a BE ensemble for each species. We also showed the zero point energy correction to the electronic BE to be around 14\% of the BE itself irrespective of the adsorbed species, and provided a scale factor to obtain the BE(0) for adsorption complexes at the amorphous ice model. This is important as it allows skipping the expensive frequency calculation by simple scaling of the electronic BEs.
The analysis of the obtained data showed that, in order to describe properly the behavior of S-bearing species, special care must be paid in the description of the dispersive interactions, which gain more importance as the number of large and heavy atoms, like sulphur, increases. When considering the amorphous ice model, which provides enough richness of sites, the resulting BE ensembles match both the experimental and the literature computed BEs. We think BE distributions rather than single BE values to be essential for providing robust parameters to be adopted in numerical models of chemical evolution in Universe.

For what concerns the sulphur depletion problem, our calculations do not solve the open questions. Nonetheless, we provide strong and relatively accurate BE values which can elucidate the features of S-bearing species in comparison with S-free molecules when adsorbed on icy mantles. An important clue is provided by the  S$_2^{\bullet\bullet}$ case in which the lowest BE implies higher diffusivity and increasing reactivity towards the formation of higher weight S-containing rings strongly bound to the ice through dispersion interactions. The CS molecule also shows some counter intuitive results when compared to the analogue CO molecule, due to a higher dipole moment and dispersion contribution. The present work also highlights the utility to compute accurately BE values to be correlated with snow lines in protoplanetary disks.

\section{Acknowledgments} 
This project has received funding within the European Union’s Horizon 2020 research and innovation programme from the European Research Council (ERC) for the projects ``Quantum Chemistry on Interstellar Grains” (Quantumgrain), grant agreement No 865657 and ``The Dawn of Organic Chemistry” (DOC), grant agreement No 741002, and from the Marie Sklodowska-Curie for the project ``Astro-Chemical Origins” (ACO), grant agreement No 811312. The Italian Space Agency for co-funding the Life in Space Project (ASI N. 2019-3-U.O), the Italian MUR (PRIN 2020, Astrochemistry beyond the second period elements, Prot. 2020AFB3FX) are also acknowledged for financial support. A.R. is indebted to the “Ramón y Cajal” program. Authors thank Gretobape for fruitful and stimulating discussions. The authors wish to thank the anonymous reviewers for their valuable suggestions.  
Supplementary Material consisting of (i) the fractional coordinates of HF-3c adsorption structure optimized for both the crystalline and amorphous ice, (ii) images of the adsorption features at crystalline periodic ice models, in which electrostatic potential maps, spin density maps (for open-shell systems) and vibrational features are displayed, (iii) a pdf file with a thorough guide to the computation of BEs and the basis sets employed for the calculations, is available at \url{https://zenodo.org/record/6798922}.

\bibliography{my_biblio}{}
\bibliographystyle{aasjournal}

\appendix
\section{Computational Details}
\label{sec:appendix}
For optimization and frequency calculations, we refer to the computational parameters adopted in the work of \cite{ferrero:2020}: the threshold parameters for the evaluation of the Coulomb and exchange bi-electronic integrals (TOLINTEG) were set to 7, 7, 7, 7 and 14 when working with crystalline ice, while for the amorphous ice to 7, 7, 7, 7 and 25, which facilitates the convergence of the self-consistent field (SCF), especially when adsorbing open-shell species.

Some species required a careful treatment to describe properly their electronic state. The use of the BROYDEN accelerator \citep{johnson:1988} and the broken-(spin)-symmetry \textit{ansatz} \citep{neese:2004,neese:2009} were necessary to obtain an optimal convergence of the SCF procedure.

For the numerical evaluation of bielectronic integrals \citep{becke:1988}, the standard pruned grid, composed of 75 radial points and a maximum of 974 angular points, was used. To sample the reciprocal space, we adopted the Pack–Monkhorst mesh \citep{pack_monkhorst}, with a shrinking factor of 2, that generates 4 \textbf{k} points in the first Brillouin zone. The only exception is for the adsorption of atomic sulphur on the crystalline ice model, which required a shrinking factor of 3, generating 5 \textbf{k} points, in order to avoid numerical noise in the results.

The Broyden–Fletcher–Goldfarb–Shanno (BFGS) algorithm \citep{broyden:1970,fletcher:1970,goldfarb:1970,shanno:1970} was used to carry out geometry optimizations.

\section{Calculation of the Binding Energies} \label{subsec:be}
When using a finite basis set of localized Gaussian functions to describe our systems, basis set superposition errors (BSSE) arise. This is the case in our BEs computed at both DFT//DFT and DFT//HF-3c.
Thus, the \textit{a posteriori} counterpoise (CP) correction of \cite{boysb:1970} was applied to compensate for this error and our BEs are therefore defined as the opposite of the CP-corrected interaction energies ($\Delta E^{CP}$), that is,

\begin{equation}
BE = -\Delta E^{CP}
\end{equation}
\begin{equation}\label{eqn:BE}
    \Delta E^{CP} = \Delta E - BSSE
\end{equation}

where the non-CP-corrected interaction energy $\Delta E$ is given by the sum of the deformation-free interaction energy ($\Delta E^*$), the deformation energy of the slab ($\delta E_S$) and the molecule ($\delta E_{\mu}$) and the lateral interaction ($E_L$) between adsorbate molecules in different replica of the cell. This quantity corresponds to the common definition of interaction energy, which in this case is the energy of the complex to which energies of the isolated species and isolated ice model are subtracted. 
\begin{multline}\label{eqn:deltaE}
    \Delta E = \Delta E^* + \delta E_S + \delta E_{\mu} + E_L = \\ E_{complex} - E_{ice} - E_{species}
\end{multline}

For the crystalline systems, inclusion of the ZPE-corrections to the BE are performed according to the equation BE(0)= BE - $\Delta$ZPE, where

\begin{equation} \label{eqn:ZPE}
   \Delta ZPE = ZPE_{complex} - ZPE_{ice} - ZPE_{species}
\end{equation}

For the amorphous case, inclusion of ZPE-corrections was done by applying a scaling factor derived from a linear correlation between BE and BE(0) in the crystalline systems, that is, BE(0) = 0.862*BE.

Finally, to check for the accuracy and refine DFT BEs for our crystalline ice model, the single- and double- electronic excitation coupled-cluster method with an added perturbative description of triple excitations (CCSD(T)) was used, in combination with a correlation consistent basis set extrapolated to the largest possible basis set.
These calculations have been performed with the Gaussian16 software package \citep{gaussian}. 
These refinements were performed through the ONIOM2 approach \citep{dapprich:1999}, dividing the systems in two parts (\textit{model} and \textit{real} systems), which are described by two different levels of theory (\textit{high} and \textit{low}). The \textit{model} system (represented by the adsorbate and the two closest water molecules) was described by the \textit{high} level of theory, CCSD(T). The \textit{real} system (that is, the whole system) was described by the corresponding DFT (\textit{low}) level of theory.

The ONIOM2-corrected BE is:

\begin{multline} \label{ccsdt}
    BE(ONIOM2) = BE(low,real) + \\ 
    BE(high,model) - BE(low,model)
\end{multline}

In this way, the $\Delta$E = BE(\textit{high,model}) - BE(\textit{low,model}) represents the correction term to the energy of the \textit{real} system due to the improved description at the \textit{high} level. This procedure ensures that if the \textit{low} level of theory is improved to arrive at the \textit{high} level of theory or if the \textit{model} system is enlarged to become the \textit{real} system, the energy of the whole system is that at the higher level of theory. 

In this work, for the calculation of the ONIOM2-corrected BEs, BE(ONIOM2), equation \ref{ccsdt} can be rewritten as

\begin{multline}
    BE(ONIOM2) = BE(DFT;\mu - ice) + \\
    BE(CCSD(T);\mu - 2\ch{H2O}) - BE(DFT;\mu - 2\ch{H2O}) 
\end{multline}

where $BE(DFT;\mu - ice)$ is the BE computed at DFT//DFT, while the BEs of the model system ($\mu$ - 2H$_2$O) are computed through single point energy calculations at CCSD(T). 

A thorough guide to the computation of BEs is available in the Supplementary Material.

\section{S-bearing vs O-bearing species} \label{subsec:geom}

\begin{figure}[!htb]
    \begin{tabular}{@{} c}
         \includegraphics[width=\columnwidth]{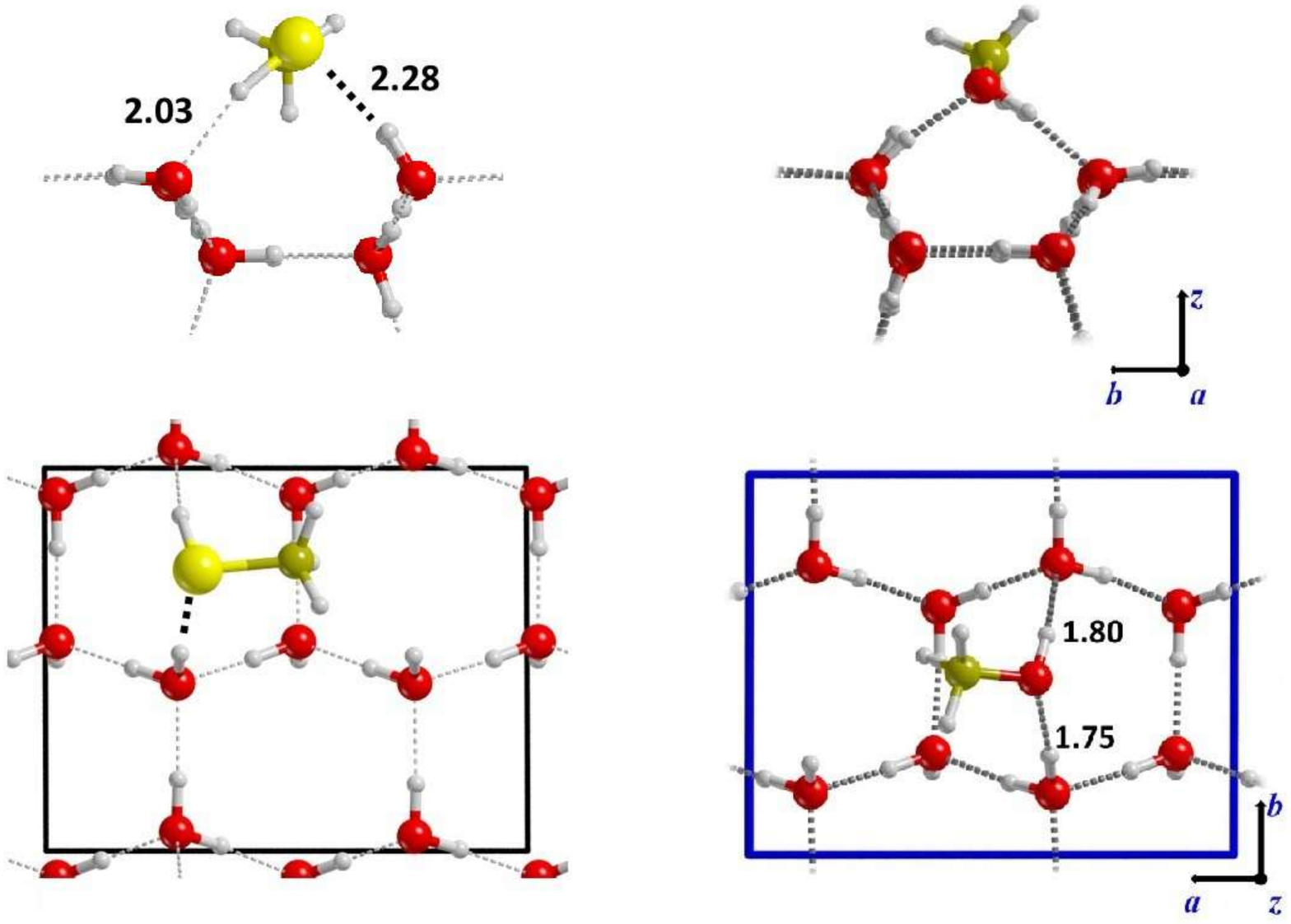} \\
         \hline
    \includegraphics[width=\columnwidth]{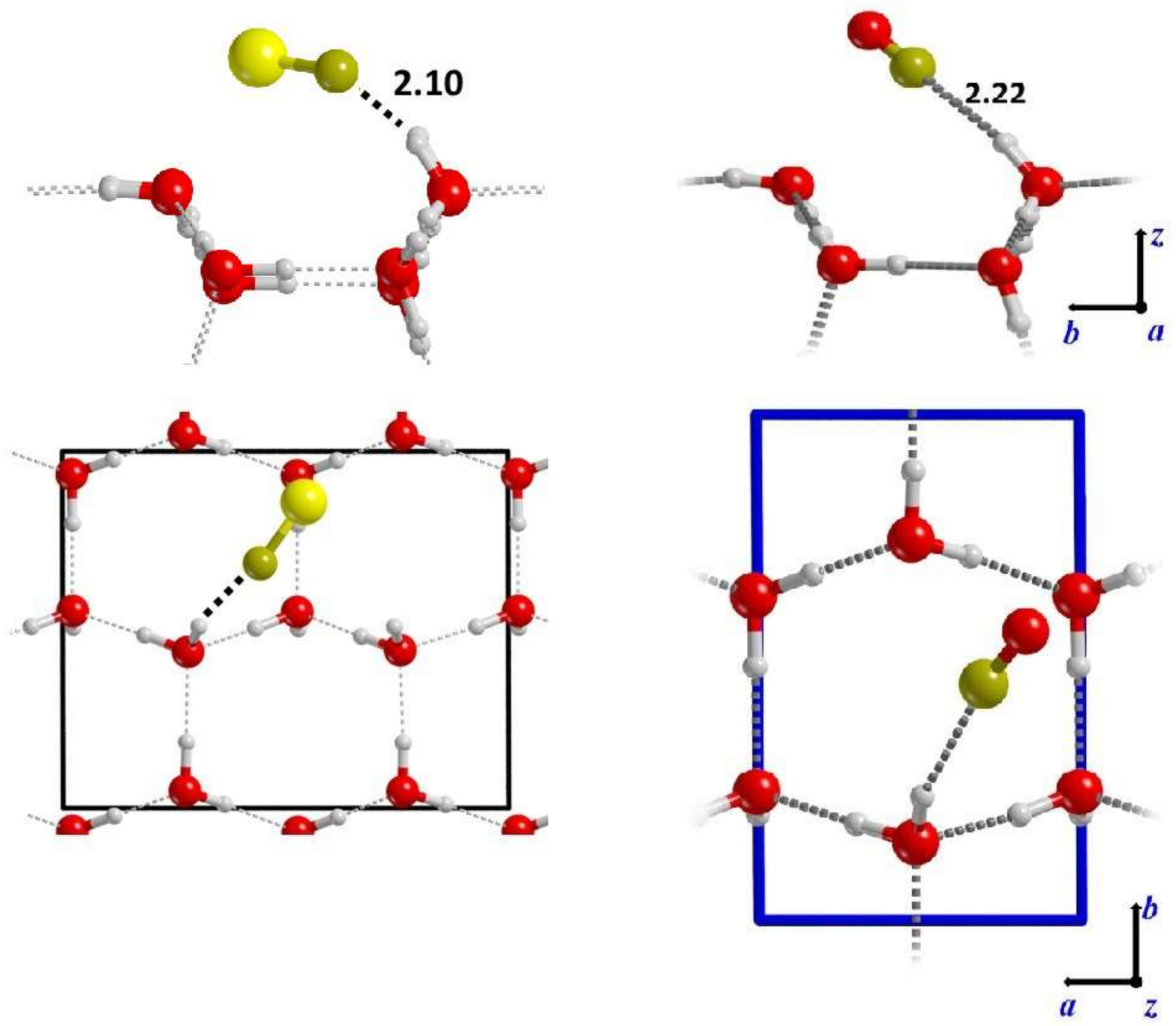} \\
    \hline
       \end{tabular}
    \caption{Adsorption geometry of S-bearing vs O-bearing species at P-ice (010) surface. Side view at the top; top view at the bottom. Complexes on the right (O-bearing species) are adapted from \cite{ferrero:2020}. Distances are given in \AA. i) CH$_3$SH versus CH$_3$OH; ii) CS versus CO.}
    \label{fig:geom}
\end{figure} 
     
\begin{figure}[!htb]
    \begin{tabular}{@{} c}
         \includegraphics[width=\columnwidth]{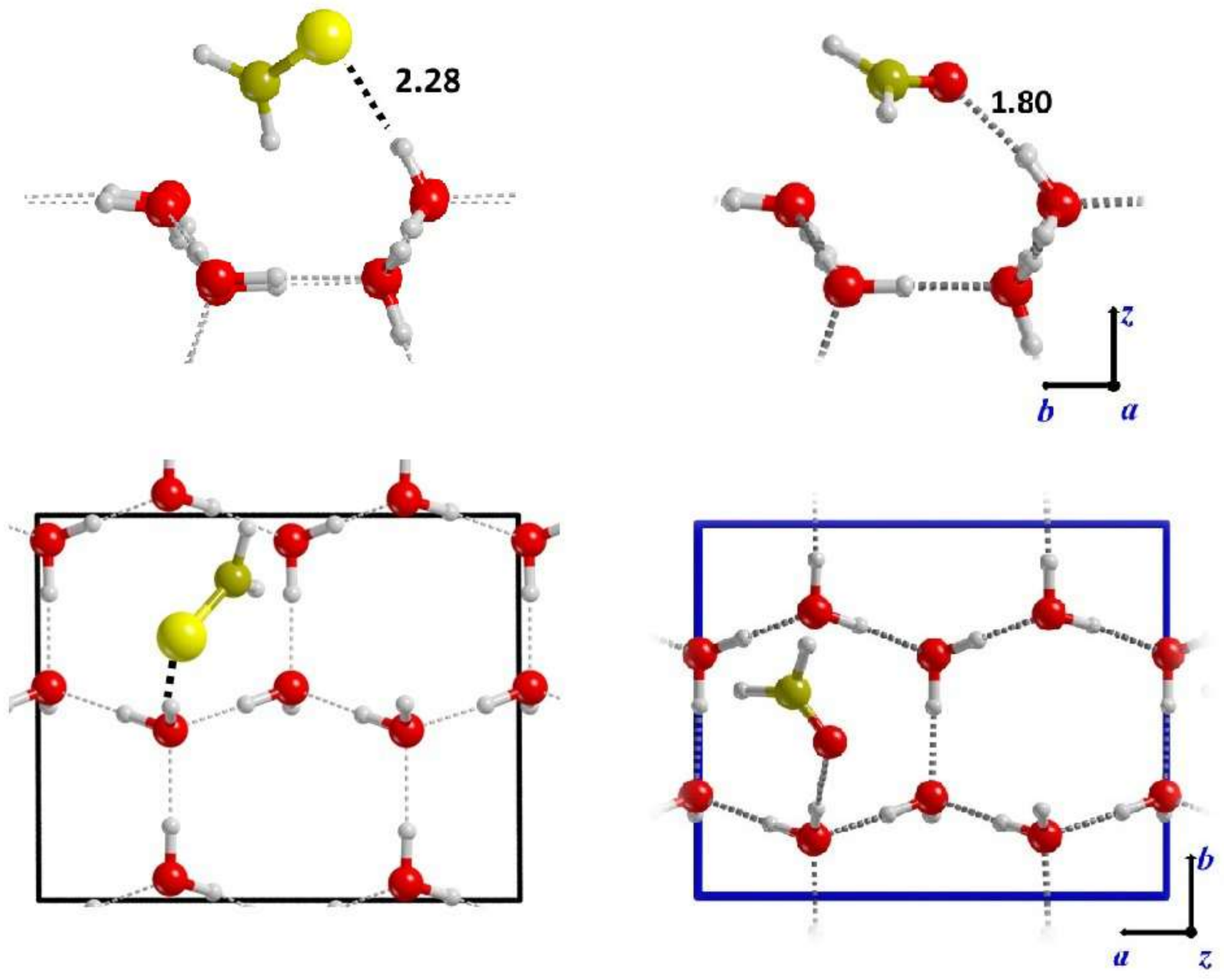} \\
    \hline
         \includegraphics[width=\columnwidth]{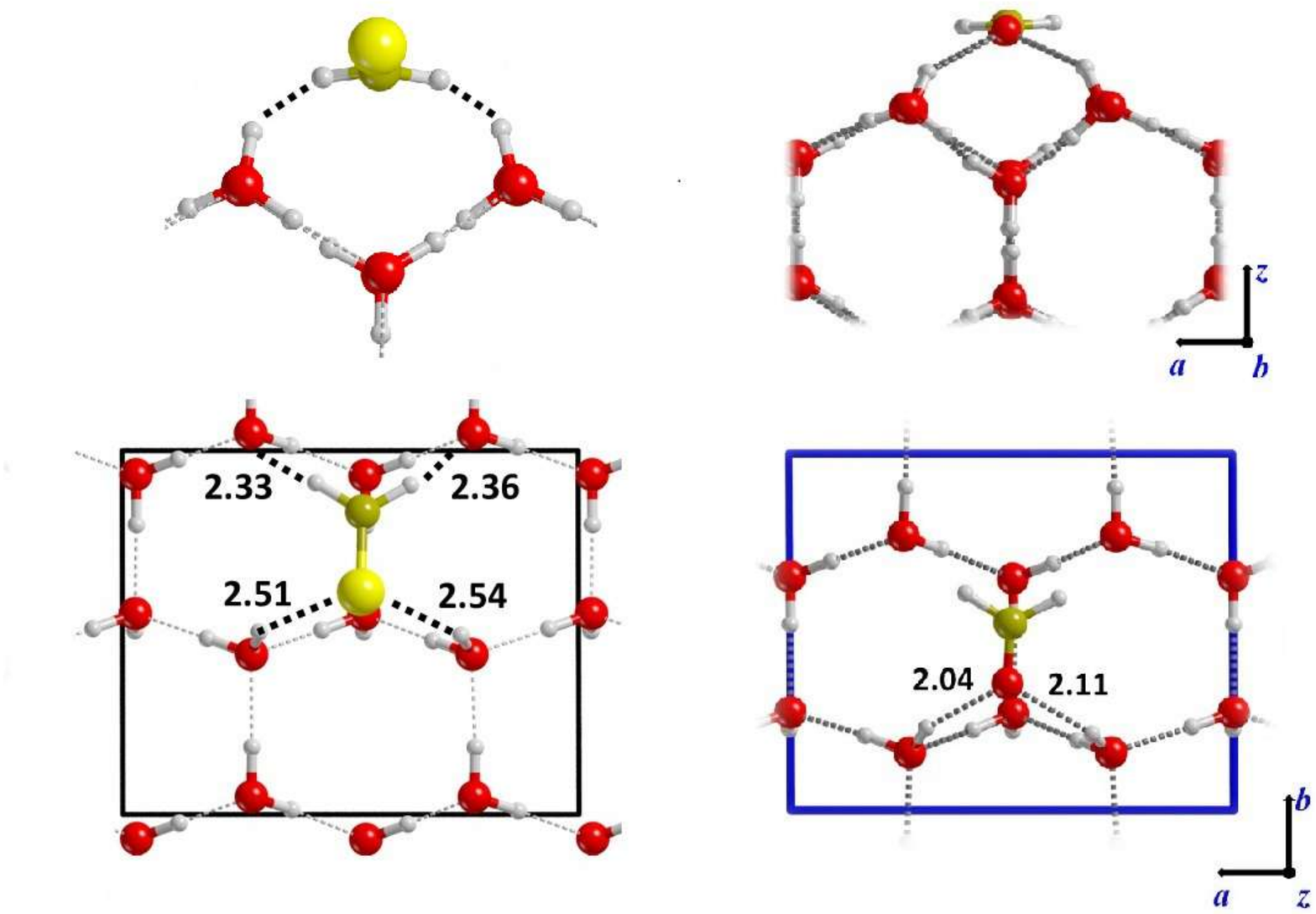} \\
    \hline
        \end{tabular}
    \caption{Adsorption geometry of S-bearing vs O-bearing species at P-ice (010) surface. Side view at the top; top view at the bottom. Adsorption complexes on the right (O-bearing species) are adapted from \cite{ferrero:2020}. Distances are given in \AA. i) First and ii) second adsorption geometry of H$_2$CS versus H$_2$CO.}
    \label{fig:geom2}
\end{figure}
    
    \begin{figure}[!htb]
    \begin{tabular}{@{} c}
         \includegraphics[width=\columnwidth]{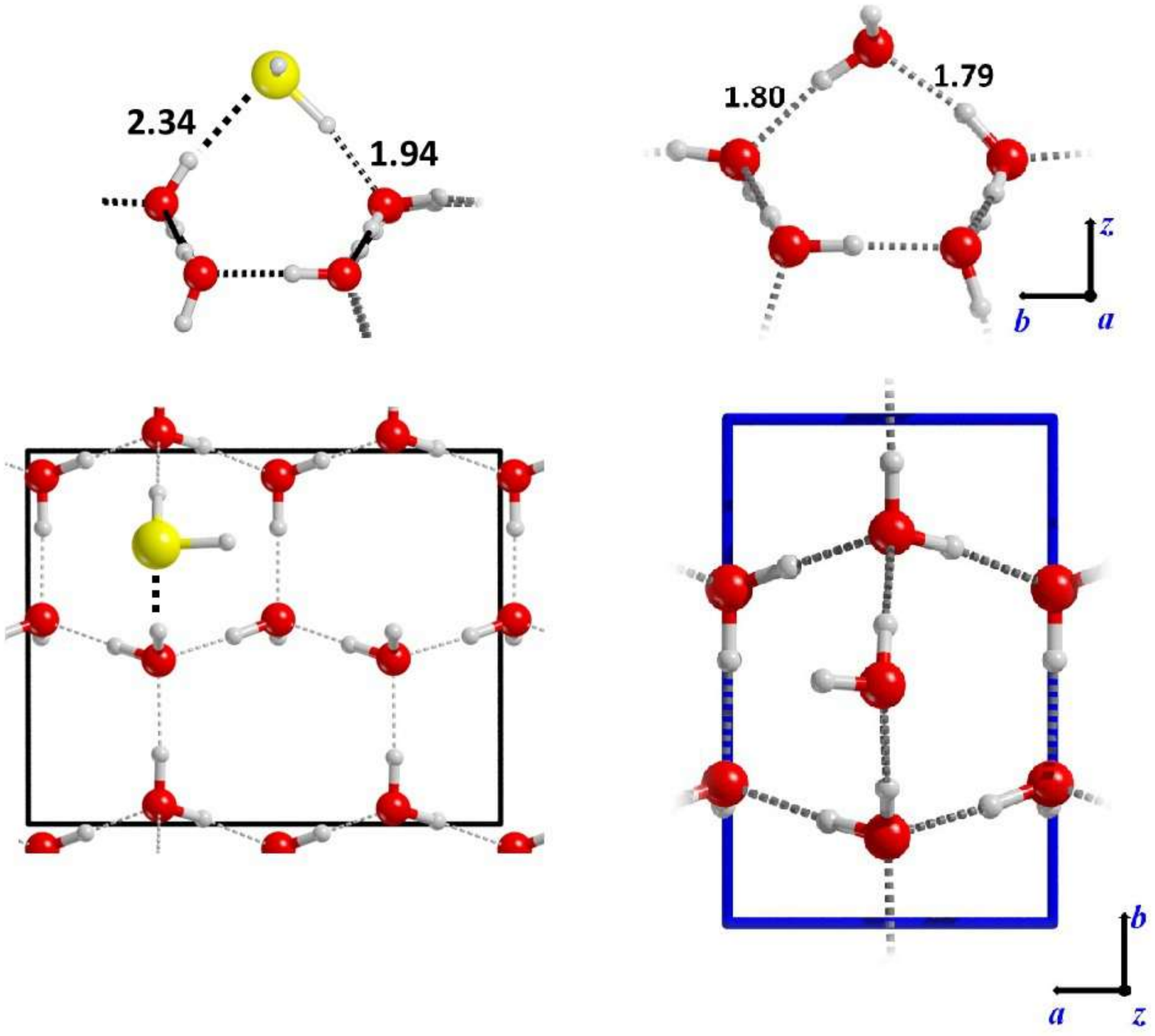} \\
    \hline
         \includegraphics[width=\columnwidth]{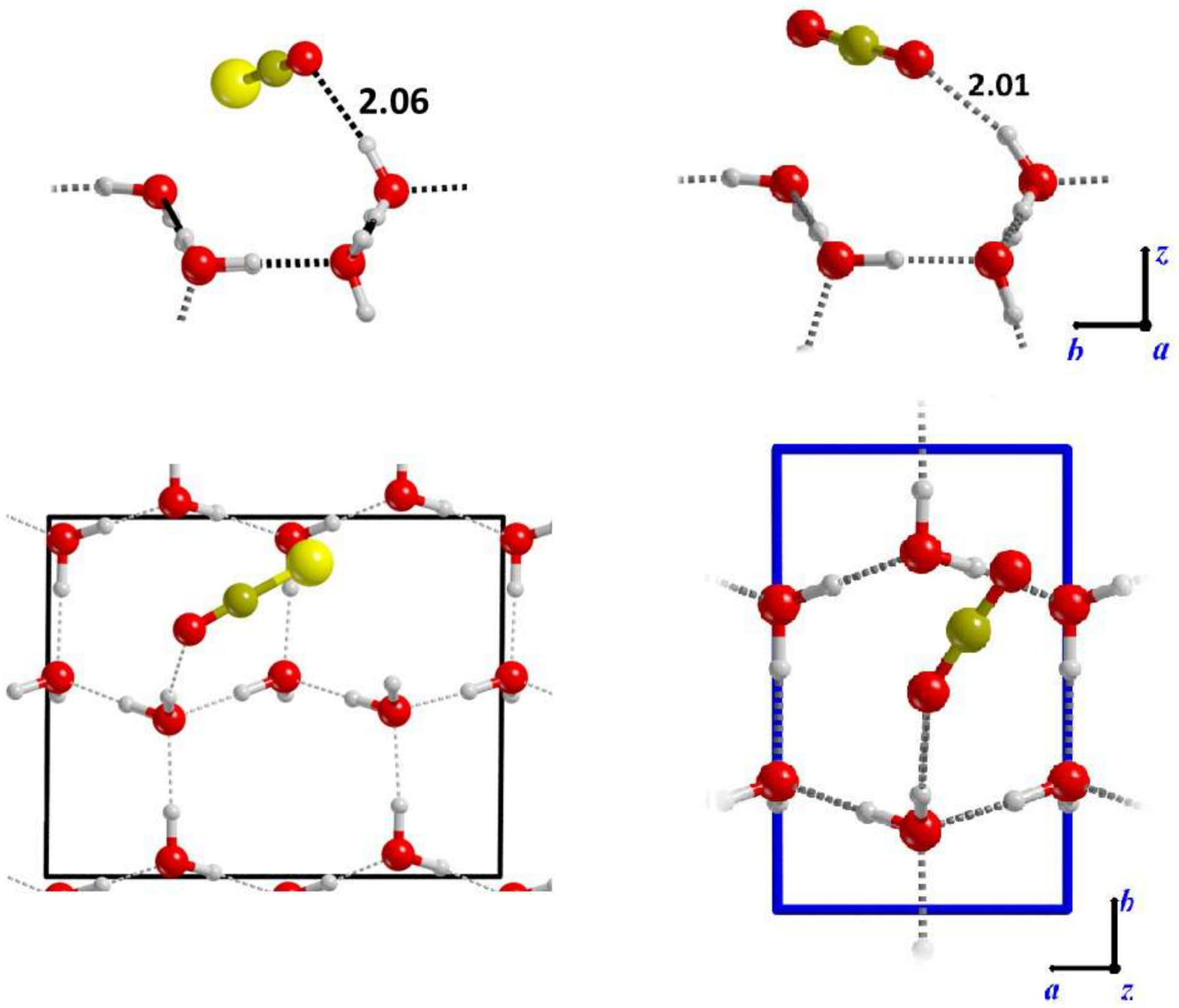} \\
    \hline
    \end{tabular}
    \caption{Adsorption geometry of S-bearing vs O-bearing species at P-ice (010) surface. Side view at the top; top view at the bottom. Adsorption complexes on the right (O-bearing species) are adapted from \cite{ferrero:2020}. Distances are given in \AA. i) H$_2$S versus H$_2$O; ii) Adsorption geometry of OCS versus CO$_2$.}
    \label{fig:geom3}
\end{figure}

\end{document}